\PassOptionsToPackage{table,svgnames}{xcolor}
\documentclass[lettersize,journal]{IEEEtran}

\usepackage{textcomp}
\usepackage{url}
\usepackage{cite}
\usepackage{verbatim}
\usepackage{stfloats} % For float placement control
\usepackage{calc}     % For calculations in lengths
\usepackage{lipsum}   % For dummy text
\usepackage{balance}
\usepackage{amsmath,amsfonts,amssymb} % Combined math packages

\usepackage{graphicx}
\usepackage{caption}
\usepackage{subcaption} % The modern package for subfigures
\usepackage{booktabs}
\usepackage{threeparttable}
\usepackage{multirow}
\usepackage{makecell}

\usepackage{soul}          % For highlighting with \hl
\usepackage{xcolor}

\definecolor{c1}{HTML}{4b5cc4}
\definecolor{c5}{HTML}{b71a3b}
\definecolor{light-gray}{gray}{0.85} % Slightly adjusted for better visibility
\definecolor{grannysmithapple}{rgb}{0.66, 0.89, 0.63}
\definecolor{green(html/cssgreen)}{rgb}{0.0, 0.5, 0.0}
\definecolor{brightmaroon}{rgb}{0.76, 0.13, 0.28}
\definecolor{codegreen}{rgb}{0,0.6,0}
\sethlcolor{white}

\usepackage[linesnumbered,lined,commentsnumbered,ruled]{algorithm2e}

\usepackage{listings}

\lstdefinestyle{interfaces}{
  float=tp,
  floatplacement=tbp,
  abovecaptionskip=0pt,
} %used in list

\usepackage{pifont}
\usepackage{bbding}
\usepackage{wasysym}
\usepackage{amssymb}
\usepackage{tikz}
\usepackage{ulem}
\usepackage{framed}
\usepackage{enumitem}

\newcommand{\var}[1]{$#1$}

\usepackage{hyperref}
\usepackage{cleveref}

\hypersetup{
    colorlinks=true,
    linkcolor=DarkBlue,
    citecolor=DarkBlue,
    urlcolor=DarkBlue,
    filecolor=DarkBlue
}

\begin{document}

\title{\textsc{Vercation}: Precise Vulnerable Open-source Software Version Identification based on Static Analysis and LLM}


\author{
    \IEEEauthorblockN{
        Yiran Cheng\IEEEauthorrefmark{1}\IEEEauthorrefmark{2}, 
        Ting Zhang\IEEEauthorrefmark{3}\IEEEauthorrefmark{5}, 
        Lwin Khin Shar\IEEEauthorrefmark{3}, 
        Shouguo Yang\IEEEauthorrefmark{4}, 
        Chaopeng Dong\IEEEauthorrefmark{1}\IEEEauthorrefmark{2}, 
        David Lo\IEEEauthorrefmark{3}, 
        Shichao Lv\IEEEauthorrefmark{1}\IEEEauthorrefmark{2}\IEEEauthorrefmark{5}, 
        Zhiqiang Shi\IEEEauthorrefmark{1}\IEEEauthorrefmark{2}, 
        Limin Sun\IEEEauthorrefmark{1}\IEEEauthorrefmark{2}
    }
    
    \IEEEauthorblockA{
    \IEEEauthorrefmark{1} Beijing Key Laboratory of IOT Information Security Technology, \\ Institute of Information Engineering, Beijing, China} 
    
    \IEEEauthorblockA{\IEEEauthorrefmark{2} School of Cyber Security, University of Chinese Academy of Sciences, Beijing, China}
    
    \IEEEauthorblockA{\IEEEauthorrefmark{3} Singapore Management University, Singapore}
    
    \IEEEauthorblockA{\IEEEauthorrefmark{4} Zhongguancun Laboratory, Beijing, China}
    
    \IEEEauthorblockA{chengyiran@iie.ac.cn, tingzhang.2019@phdcs.smu.edu.sg, lkshar@smu.edu.sg, \{yangshouguo, dongchaopeng\}@iie.ac.cn, davidlo@smu.edu.sg, \{lvshichao, shizhiqiang, sunlimin\}@iie.ac.cn}
}

\markboth{Journal of \LaTeX\ Class Files,~Vol.~14, No.~8, August~2025}%
{Shell \MakeLowercase{\textit{et al.}}: A Sample Article Using IEEEtran.cls for IEEE Journals}


\maketitle

\renewcommand{\thefootnote}{} %将脚注符号设置为fnsymbol类型，即特殊符号表示
\footnotetext{\IEEEauthorrefmark{5} Shichao Lv and Ting Zhang are the corresponding authors.} %对应脚注[1]

\begin{abstract}
Open-source software (OSS) has experienced a surge in popularity, attributed to its collaborative development model and cost-effective nature. However, the adoption of specific software versions in development projects may introduce security risks when these versions bring along vulnerabilities.
  %In light of this, having precise knowledge of the specific vulnerable versions within OSS becomes imperative for effectively mitigating potential security threats. 
  Current methods of identifying vulnerable versions typically analyze and extract the code features involved in vulnerability patches using static analysis with pre-defined rules. They then use code clone detection to identify the vulnerable versions.
  These methods are hindered by imprecision due to (1) the exclusion of vulnerability-irrelevant code in the analysis and (2) the inadequacy of code clone detection.
  %Additionally, relying solely on syntactic-level code features to identify vulnerable versions has proven to be a significant limitation in this task.
  This paper presents \textsc{Vercation}, an approach designed to identify vulnerable versions of OSS written in C/C++. 
  \textsc{Vercation} combines program slicing with a Large Language Model (LLM) to identify vulnerability-relevant code from vulnerability patches. It then backtracks historical commits to gather previous modifications of identified vulnerability-relevant code. We propose code clone detection based on expanded and normalized ASTs to compare the differences between pre-modification and post-modification code, thereby locating the vulnerability-introducing commit (\textit{vic}) and enabling the identification of the vulnerable versions between the vulnerability-fixing commit and the \textit{vic}. %We then identify the versions affected by a vulnerability based on the vulnerability-fixing commit and the corresponding \textit{vic}.
  % This paper presents \textsc{Vercation}, an approach designed to identify vulnerable versions of OSS written in C/C++. 
  % By combining program slicing with a Large Language Model (LLM), 
  % we are able to improve the precision in terms of 
  % reasoning about the vulnerability and 
  % identifying vulnerability-relevant code, given a vulnerability patch.
  % We then backtrack historical commits to gather previous modifications of identified vulnerability-relevant code. 
  % We propose semantic-level code clone detection to compare the differences between pre-modification and post-modification code, thereby locating the vulnerability-introducing commit (\textit{vic}). We then identify the versions affected by a vulnerability based on the vulnerability-fixing commit and the corresponding \textit{vic}.
%With the help of security patches, we generate vulnerability signatures through an analysis of vulnerability behavior. Subsequently, a commit backtrack of the vulnerability signatures is conducted to pinpoint the vulnerability-introducing commit.
%The commit traceback mechanism relies on semantic invariance as a criterion for continuing the traceback.
  We curate a dataset linking 122 OSS vulnerabilities and 1,211 versions to evaluate \textsc{Vercation}. On this dataset, our approach achieves an F1 score of 93.1\%, outperforming current state-of-the-art methods. More importantly, \textsc{Vercation} detected 202 incorrect vulnerable OSS versions in NVD reports. 
\end{abstract}


\section{Introduction}
%Nowadays, open-source software plays a vital role in software development by providing pre-existing, reliable, and tested solutions to common programming challenges, enabling developers to focus more on specific project requirements and less on reinventing fundamental functionalities. 
Open-source software (OSS) has become increasingly popular in recent years, thanks to its collaboration and cost-effectiveness. %transparent development process. 
In the rapidly evolving world of OSS, numerous versions exist due to continuous evolution.
%According to a recent survey conducted by the software development platform Stack Overflow, approximately 87\% of developers reported using open-source software in their projects~\cite{stackoverflow_survey}. 
While OSS plays a pivotal role in expediting software development, the integration of particular software versions in development projects can pose security risks, as these versions may contain vulnerabilities. 
%it can also pose vulnerabilities that are specific to certain versions. 
Therefore, having a comprehensive knowledge of vulnerable versions of OSS becomes imperative for software developers. 
%it is crucial for software developers to be informed with a comprehensive knowledge of vulnerable versions of OSS.
%it is crucial to identify vulnerable versions accurately.
%For example, In December 2021, a critical vulnerability was discovered in Apache Log4j, a popular open-source logging framework used by many organizations. This vulnerability allows attackers to remotely execute code on vulnerable systems using the Log4j library~\cite{log4j}. 
%Due to its widespread impact and potential for exploitation, the vulnerability was rated critical.
%As security vulnerabilities exist in specific versions, it is crucial to identify vulnerable versions accurately. 
%When OSS developers have a thorough understanding of which versions are affected, they can concentrate their efforts on releasing patch versions that address these issues specifically. This also makes it easier for maintainers of projects that depend on OSS to quickly determine whether their project is affected by a vulnerability, allowing them to update their software versions accordingly.

%Public vulnerability databases serve as repositories that collect and organize reports detailing vulnerabilities in software products. These databases then disseminate information regarding the affected versions of the software. 
Public vulnerability repositories collect vulnerability reports of software products and disseminate information regarding the affected versions of the software.
The National Vulnerability Database (NVD)~\cite{NVD}, recognized as the largest public vulnerability database, employs the Common Platform Enumeration (CPE) format to store information about vulnerable versions. However, 
%lacking complete information and targeted analysis, 
the NVD often encompasses all versions before the reported vulnerability or designates only the versions mentioned in the report as vulnerable. For instance, CVE-2018-5785 reported only version v2.3.0 as vulnerable in CPE~\cite{CVE-2018-5785}. Yet, upon manual validation, it was uncovered that versions v2.1.1 to v2.3.0 are all susceptible to the vulnerability. 
Recent research~\cite{incomplete_dong2019towards, incomplete_mu2018understanding, incomplete_tan2021locating, nguyen2013reliability} indicates that incomplete and incorrect information about vulnerable versions is prevalent in such reports. 
Therefore, there is a need for an automated method that can more accurately identify vulnerable versions of released OSS vulnerabilities.

\vspace{4px}
\noindent\textbf{Two limitations of current approaches.} 
The approaches for confirming vulnerabilities involve utilizing Proof of Concept (PoC) to trigger the vulnerability.
PoC triggers vulnerabilities through dynamic execution, offering conclusive evidence of their existence.
However, executing a PoC for each software version is a time-intensive process that requires meticulous environment setup. Additionally, the %applicability of the 
PoC input may not be universal across all vulnerable versions~\cite{dai2021facilitating}. 
Therefore, current researches use static analysis to identify vulnerable versions~\cite{da2016framework, rosa2021evaluating, nguyen2016automatic, rodriguez2020bugs, xiao2020mvp, shi2022precise, jang2012redebug, woo2022movery}. 
These methods typically consist of two steps: extracting vulnerability features from code snippets and identifying vulnerable versions through code clone detection of these vulnerability features.
%The effectiveness of vulnerability feature extraction heavily relies on human expertise in formulating vulnerability rules.
%set by static analysis tools.
Existing methods typically extract vulnerability features using static analysis in which vulnerability patterns are pre-defined by human experts: some consider data and control dependencies from the patched code~\mbox{\cite{xiao2020mvp,shi2022precise}}; some consider using hashes of entire functions~\mbox{\cite{woo2021v0finder}}; and some consider only the patch code~\mbox{\cite{jang2012redebug, bao2022v}}.
\hl{These methods generally produce many false positives or false negatives because of the inherently imprecise nature of static analysis and the incompleteness of pre-defined vulnerability patterns, making it difficult to distinguish vulnerability-relevant code from irrelevant code} (\noindent\ding{182}).
%Existing methods for vulnerable code clone detection are limited to syntactic signatures (such as Levenshtein algorithm~\cite{bao2022v} and hash value comparison~\cite{xiao2020mvp, woo2021v0finder}), without taking into account code changes that still preserve the semantics of functions (\ding{183}).
The majority of the methods for code clone detection are function-level approaches~\cite{fang2020functional} whereas real-world vulnerability logic is typically confined to (often a few) statements only. There are statement-level code clone detection methods but they are limited to textual signatures (such as Levenshtein algorithm~\cite{bao2022v} and hash value comparison~\cite{xiao2020mvp, woo2021v0finder}), without taking into account the code changes that reorganize code while preserving its functionality. These limitations hinder the effectiveness of vulnerability detection in practical scenarios (\ding{183}).

These limitations primarily stem from the lack of reasoning about the context of the vulnerable code being analyzed. %Currently, this reasoning effort is done via manual analysis, which is inefficient and not scalable. 
Recently, large language models (LLMs) have shown remarkable performance in various code-related tasks, including code generation~\cite{du2024evaluating,li2023large} and code summarization~\cite{ahmed2024automatic,ahmed2022few}. This strong performance demonstrates LLMs' capability to recognize code patterns and correlations at both syntactic and semantic levels.
%in code-related tasks including code generation~\cite{du2024evaluating,li2023large} and code summarization~\cite{ahmed2024automatic,ahmed2022few}.
As such, we aim to address the first limitation by leveraging the power of LLMs. Essentially, we hypothesize that LLMs can be leveraged to address the imprecision of static analysis and the incompleteness of human-defined vulnerability patterns (\noindent\ding{182}).
%Therefore, to address existing methods' limitations, in this work, we explore the use of LLM to assist with vulnerability comprehension task.  
%and semantic-level code clone detection in the context of vulnerability versions identification.
%\noindent\textbf{Challenges.} 
However, there is a challenge of \emph{Prompt engineering} in leveraging an LLM in our problem domain. The design of prompts is pivotal in directing the LLM to produce desired responses.
%exerting a significant impact on their performance. 
In vulnerability comprehension tasks, well-crafted prompts should encompass the code for analysis and offer a lucid description of the analysis objectives. However, in the case of current LLMs, merely providing all the vulnerable function codes and directly instructing vulnerability analysis often yields suboptimal outcomes. 
%Meanwhile, currently, most LLMs can only accept a limited amount of context (GPT-4.0 supports up to 32,000 tokens). 
When the function code is too long, the LLM's ability to understand the relationships between distant parts of the context may diminish~\cite{li2024longcontext, song2024hierarchical}.
Hence, formulating prompts to guide LLMs in generating desired vulnerability logic analyses poses a challenge.

%\emph{Semantic-level code clone detection}: 
Due to variations in programming styles, variable naming, and structural differences, code clones may manifest different changes. Furthermore, developers often streamline and refactor code through method outlining. Such code structural changes often mislead code clone detection-based approaches to incorrectly identify the vulnerable versions. %Consequently, accurately detecting semantic-level clones remains a challenge. 
%Furthermore, from our manual analyses of several code repositories, we have observed that the majority of the vulnerability code fixes are at the statement level, while existing code clone approaches are focused on function-level detection.
\hl{Furthermore, prior studies}~\cite{hin2022linevd,duan2019vulsniper} \hl{found that vulnerabilities can often be localized to a few key lines, making file- and function-level vulnerability detection overly coarse-grained.}
The limitation motivated us to propose a statement-level code clone detection approach to address the second limitation of current approaches (\noindent\ding{183}).

\vspace{4px}
\noindent\textbf{Our approach.} We propose \textsc{Vercation} (\underline{V}ulnerable v\underline{er}sion identifi\underline{cation}), a novel method to identify vulnerable versions of open-source C/C\texttt{++} software utilizing a symbiotic combination of static analysis, LLM, and code clone detection. Given a vulnerability fixing commit (\textit{vfc}), \textsc{Vercation} applies program slicing to extract vulnerability-related statements as vulnerability features, leverages the capability of LLM in code understanding to refine the extracted features, and performs semantic-level clone detection on vulnerability features in code changes. This hybrid method effectively overcomes the practical limitations of existing approaches.
More specifically, \textsc{Vercation} automatically preprocess the fixing commits and construct prompts based on the Few-shot and Chain-of-Thought (CoT) strategies, enabling the LLM to reason with the vulnerability and identify the most probable vulnerable statements as features. 
Subsequently, \textsc{Vercation} traces earlier modifications of vulnerability features and applies a clone detection method based on expanded Abstract Syntax Trees (ASTs) to pinpoint the vulnerability-introducing commit \textit{vic}. 
%The proposed semantic clone detection method is based on expanded and normalized Abstract Syntex Trees (ASTs), applied to vulnerable lines before and after modifications (as detailed in Section 3.2). 
Finally, \textsc{Vercation} identifies vulnerable versions between the \textit{vic} and \textit{vfc}. %(as detailed in Section 3.3).

\noindent\textbf{Evaluation.}
We meticulously curated a ground-truth dataset for evaluation, encompassing 12 commonly used OSS projects, 122 Common Vulnerabilities and Exposures (CVEs), and a total of 1,211 OSS versions. 
This dataset comprised every patch released through Git with respect to those 122 CVEs.
The first author engaged in manual vulnerability validation with the help of public Proof-of-Concept (PoC), meticulously labeling the presence of vulnerabilities across software versions. On this dataset,
\textsc{Vercation} demonstrated both higher precision (91.8\%) and recall (94.5\%) compared to state-of-the-art methods (SOTAs), including V-SZZ~\cite{bao2022v}, V0Finder~\cite{woo2021v0finder}, V1SCAN~\cite{clone_woo2023v1scan}, VERJava~\cite{sun2022verjava} and Vision~\cite{wu2024vision}.
We conducted an ablation study using three different LLMs (GPT-4~\cite{gpt4}, CodeLlama~\cite{roziere2023codellama}, and DeepSeek-V3~\cite{deepseekai2024deepseekv3technicalreport}) to evaluate their vulnerability comprehension capabilities. Utilizing the Few-shot and CoT combined strategy, DeepSeek-V3 achieved an F1 score of 93.1\%,
%, for the vulnerable version identification task
significantly improving the F1 score of Joern parser~\cite{Joern} by 92.8\%, a commonly-used static analysis tool. More importantly, during the evaluation, we found 202 version errors in the NVD reports. \textsc{Vercation} has also been shown to be efficient, analyzing each vulnerability on an average of 28.61 seconds.
% We created a ground-truth dataset for evaluation, which comprised 11 OSS, 74 CVEs and 1013 OSS versions. This dataset consisted of all the patches released through Git. To ensure accuracy, we manually examined and labeled the presence of vulnerabilities across 10,476 software-version pairs. 
% In our experiments, \textsc{Vercation} achieves 93\% precision and 98\% recall, significantly outperforming existing techniques V-SZZ~\cite{bao2022v} and V0Finder~\cite{woo2021v0finder} for identifying vulnerable versions. Moreover, our evaluation has revealed a notable error rate in the official National Vulnerability Database (NVD) regarding information on released vulnerable versions.  
% In terms of efficiency, \textsc{Vercation} takes an average of 28.61 seconds to analyze each vulnerability.\\[3pt]

\noindent\textbf{Contributions.} The main contributions of this paper are as follows:
\begin{itemize}[leftmargin=*]
	% \item We constructed a comprehensive dataset comprising 10,476 unique vulnerability-version pairs from 1,013 versions of 74 OSS CVE vulnerabilities across 11 OSS projects, which were labeled through a combination of PoC input validation and manual verification.
 %We constructed a comprehensive dataset containing 74 OSS CVE vulnerabilities with 1013 versions across 11 OSS projects. The 10,476 unique vulnerability-version pairs are labeled through a combination of PoC input validation and manual verification.
              %We constructed a comprehensive dataset containing 74 OSS CVE vulnerabilities and 1013 OSS versions (10476 vulnerability-version pairs), utilizing a of PoC input validation and manual confirmation. This dataset covers vulnerabilities from 11 OSS projects.
	\item %We present \textsc{Vercation}, an automated vulnerable version identification approach powered by an LLM. 
    Unlike previous efforts that heavily relied on pre-defined patterns of static analysis tools, we present \textsc{Vercation}, \hl{the first framework to integrate the reasoning capability of LLM for vulnerable version identification tasks,} 
 %effectively guiding LLMs to generate vulnerability logic 
 through the use of a multi-strategy universal prompt engineering.
 %include signature generation by analyzing vulnerability behavior and commit backtrack based on the semantic invariance of vulnerability signatures.
	% \item We implement a prototype of our approach and report its performance in our dataset, which provides high accuracy with 93\% precision and 98\% recall and outperforms existing techniques. Considering the widely existing missing vulnerable version reported in NVD, \textsc{Vercation} can significantly enhance the completeness of official vulnerability reports by recognizing the missing vulnerable versions.
 \item \textsc{Vercation} presents \hl{a solution based on expanded and normalized AST to address the structural modifications in the clone detection challenge, which was designed for refactoring commit identification during vulnerability backtrack.} 
  \item We curated a dataset including 122 published CVEs containing 1,211 versions. This extensive dataset was curated across 12 OSS projects and underwent meticulous labeling through a combination of PoC input validation and manual verification.
 \item We have implemented a prototype of our approach and assessed its performance using our dataset, achieving the F1 score of 93.1\%, improving SOTAs by 8.1\% to 108.7\%. We also evaluated three contrasting LLMs --- a commercial, closed model (GPT-4) and open models (CodeLlama and DeepSeek-V3) --- in our approach and reported that they achieved similar performances in our context when appropriate prompting strategies were applied. More importantly, by applying our approach, we have detected 202 incorrect vulnerable OSS versions in NVD reports. %Therefore, \textsc{Vercation} offers a practically useful solution to complement NVD  
 %Notably, in light of the prevalent issue of missing vulnerable versions in the National Vulnerability Database (NVD) reports, 
 %\textsc{Vercation} offers a practically useful solution demonstrated by reporting a significant number of version errors in the NVD reports in our experiments. 
 %It has the potential to substantially augment the completeness of official vulnerability reports by adeptly recognizing and addressing these missing vulnerable versions.
 %The result shows that the vulnerability version information provided by NVD has many false negatives, We show a significant delay in the vulnerability introduce-to-fix and discovery-to-fix.
\end{itemize}

The source code of \textsc{Vercation} and our curated dataset are publicly available at \url{https://github.com/Veronica-L/Vercation}.

%CVE信息不可靠，利用Poc无法实现大规模

%现有方法的缺陷：
%V-SZZ 1.只关注补丁，把补丁作为漏洞的引入点 2.对于只增加了行的补丁，方法无用
%Precise论文 1.针对PHP，无法应用在C/C++上 2.设定了几个关键函数，不一定适用于C/C++

%summarize challenge
%1. 相关行的确认
%2. 代码相似性问题（语法&语义）
%3. 从inducing commit到vulnerability-fixing commit的精确版本范围确认

%our approach and result

%contribution

\section{Background and Motivation}
In this section, we clarify the target problem, introduce the LLM and discuss the motivation with two examples.

\subsection{Problem Statement}
Commits serve as comprehensive records of OSS development, functioning as vital checkpoints for tracking the chronological evolution of code changes. 
They enable developers to revisit specific points on a particular date or time.
%本文研究了TPL漏洞的受影响版本分析问题。这个问题可以表述如下。在一个TPL的开发中，某些commit会引入漏洞被称为VIC，之后在vulnerability-fixing commit中漏洞被修复，The (un)affected version analysis aims to assess which versions are affected in VIC - PC.
This paper centers on vulnerable version identification in OSS vulnerabilities. Within the development of an OSS, certain commits introduce vulnerabilities, referred to as vulnerability-introducing commits (\textit{vic}), which are later fixed in vulnerability fixing commits (\textit{vfc}). The vulnerable version analysis aims to pinpoint the \textit{vic}, allowing us to assess which versions are susceptible to the vulnerabilities.
%In this paper, we focused on the vulnerable version identification of the OSS vulnerability. During the development of an OSS, some commits introduce vulnerabilities (vulnerability introducing commit, e.g., \textit{vic}), which are later fixed in the vulnerability-fixing commit (\textit{pc}). The aim of vulnerable version analysis is to locate the \textit{vic} so that we can assess which versions are affected by the vulnerabilities.
%We suppose that the initial function is introduced vulnerability in $vic_1$, $f_{vd}$ is the function after refactoring commit, which has a similar code syntax with $f_v$, $f_p$ is the patched function in \textit{pc} (see Figure~\ref{}).

%We assume that the initial function is modified by \textit{vic} and becomes a vulnerable function $F_v$. Then the refactoring commits optimize the function code and update it to function $F_{vr}$. Finally, software analysts discover the vulnerability and patch the code in the vulnerability-fixing commit, resulting in the final function $F_p$. An example timeline is depicted in Figure~\ref{fig:timeline}. It is important to note that the refactoring commit only optimizes the code syntax without altering the code's functionality and semantics, so $F_{vr}$ remains vulnerable as well.

\begin{figure}[h]
	\centerline{\includegraphics[width=0.50\textwidth]{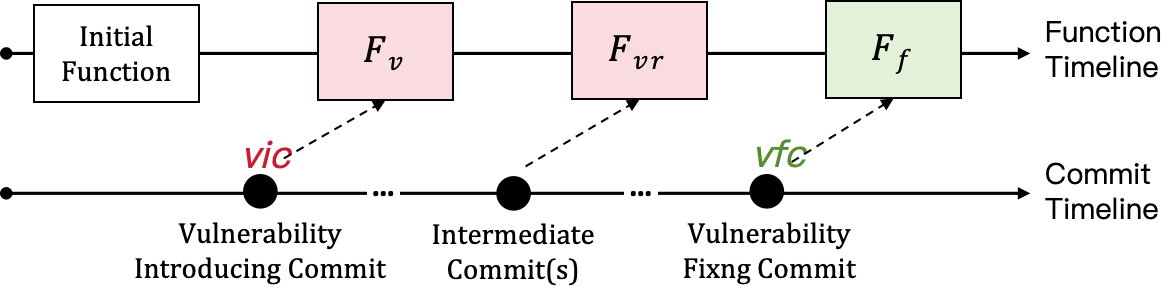}}
	\caption{Function Timeline from Vulnerability Introduction to Vulnerability Fixing.} 
    \label{fig:timeline}
\end{figure}

We posit that the initial function undergoes modification by a \textit{vic} and transforms into a vulnerable function $F_v$. There probably exists some subsequent commits such as feature-adding commits and refactoring commits to optimize the function code, which is denoted as $F_{vr}$. Ultimately, software analysts discover the vulnerability and fix the code in the \textit{vfc}, resulting in the final function $F_f$.
An illustrative timeline is presented in Figure~\ref{fig:timeline}. 

%It is necessary to emphasize that the refactoring commit exclusively optimizes the code syntax without altering the functionality and semantics of the code. Consequently, $F_{vr}$ remains vulnerable as well. \hl{Refactoring commits are common in OSS development, however, they are not necessarily contained in the function timeline.}

% \subsection{Code Syntax and Semantics}
% Program code can be analyzed from two fundamental perspectives: syntax and semantics~\cite{ma2024unveiling, fang2020functional}. Code syntax refers to the structural rules and patterns of the programming language, including how tokens, expressions, and statements are organized. It focuses on the form and structure of the code, such as variable declarations, function definitions, and control flow constructs. In contrast, code semantics refers to the actual meaning or behavior of the code when executed. Two code snippets with different syntactic structures may have the same semantic meaning. For example, the two code snippets \texttt{if (size > max) \{size = max;\}} and \texttt{size = min(size, max);} have different syntactic structures but achieve the same semantic goal of ensuring that \texttt{size} does not exceed \texttt{max}. Code refactoring often changes syntax while preserving semantics. Traditional code clone detection approaches that rely solely on syntactic matching may fail to identify semantically equivalent code after refactoring, leading to missed identifications.

\subsection{Large Language Model}
Large Language Models (LLMs) have revolutionized natural language processing and have demonstrated remarkable capabilities in various code-related tasks. 
These models, extensively trained on vast amounts of text and code repositories, can understand and generate human-like text and code.
Recent advancements in LLMs, such as GPT-3~\cite{gpt3}, GPT-4~\cite{gpt4}, DeepSeek~\cite{guo2024deepseek} and CodeLlama~\cite{roziere2023codellama}, have demonstrated impressive performance in code generation~\cite{du2024evaluating, li2023large}, code summarization~\cite{ahmed2024automatic,ahmed2022few} and program repair~\cite{fan2023programrepair, hossain2024programrepair}.

The success of LLMs in code-related tasks can be attributed to their ability to capture complex patterns and contextual relationships in code~\cite{wang2023codet5+,roziere2023codellama}. 
Through Pattern Recognition, LLMs learn correlations and patterns between code snippets from large-scale training data, enabling them to identify code with different syntax but similar functionality. Additionally, LLMs can infer the purpose of code snippets based on their context.
This makes them particularly suitable for tasks that require a deep understanding of code structure and functionality like vulnerability detection. 
However, the effectiveness of LLMs heavily depends on the quality of prompts used to guide their responses. Prompt engineering, the process of designing effective prompts, has become a crucial area of research in leveraging LLMs for specific tasks~\cite{zhou2022prompt,chen2023prompt}.

Despite their reasoning capabilities, LLMs also face challenges in code analysis tasks such as handling very long code sequences. Therefore, in our work, instead of blindly applying LLM as a standalone tool, we first use static analysis to extract candidate vulnerability-related codes and then leverage the semantic capturing capabilities of LLMs to improve the accuracy of extracted codes. %a novel application in the field of software security analysis.

% Prompts are used as user-provided input to interact with LLMs to accomplish downstream tasks. The quality and relevance of the response generated by the LLM heavily depend on the prompt's quality. 
% %Therefore, prompt engineering and prompt tuning are crucial for our task.
% Prompt engineering is a crucial process of carefully designing the prompts to generate a specific output. The different ways of writing prompts can make a lot of difference in the output generated~\cite{zhou2022prompt,chen2023prompt}.
% %Prompt tuning is an efficient, low-cost way of adapting an LLM to new downstream tasks without retraining the model and updating its weights.

\subsection{Motivating Examples}
\label{sec:motivating}
%We presented two examples of disclosed vulnerable code in different scenarios. Now all the issues have been addressed and fixed by the development team. The main points we want to highlight here are
We present two examples of disclosed vulnerable code in two distinct scenarios, which motivate our work. Both of these vulnerabilities have now been effectively resolved by the development team. The key points we wish to underscore are as follows: 
1) The significance of understanding vulnerability behavior in the process of discovering \textit{vic}.
2) The challenge of accurately pinpointing \textit{vic} due to the code structural modifications introduced by the intermediate commits.
%1) The importance of comprehending vulnerability behavior in discovering \textit{vic}. 2) Due to the syntax changes introduced by the refactoring commit, \textit{vic} is difficult to accurately locate.

% \textbf{Example 1)} We use a vulnerability in OpenSSL as a running example to illustrate the limitations of existing approaches further in vulnerability logic discovery and to motivate the idea of \textsc{Vercation}. OpenSSL is an open-source software that provides cryptographic functionality including implementations of SSL and TLS protocols~\cite{openssl}. Listing~\ref{CVE-2021-23840} shows the patch for CVE-2021-23840, which fixes an integer overflow vulnerability in function \texttt{evp\_EncryptDecryptUpdate}.
% The patch adds a sanitizing check at Line 13-17 for the multiple of the block length \texttt{(inl-j)\&\~(bl-1)} to ensure that this amount of data, plus the one block that processed from \texttt{ctx->buf} does not exceed \texttt{INT\_MAX}, which is used as parameters in function \texttt{memcpy}.
% A remote attacker can make the input length close to \texttt{INT\_MAX}, resulting in an integer overflow on Line 18, which could cause the application to behave incorrectly or crash.

\textbf{Example 1)} A vulnerability in \texttt{FFmpeg} (CVE-2017-14169~\cite{CVE-2017-14169}) shows the limitation of existing approaches in terms of capturing vulnerability logic. %and to motivate the idea of \textsc{Vercation}. 
%\texttt{FFmpeg} is an open-source software designed for multimedia processing. 
The code snippet of \textit{vfc} is shown in Listing~\ref{list:CVE-2017-14169}. 
%addressing an integer signedness error.
A sanitizing check was added at Line 13 for the \texttt{item\_num}. %, preventing it from turning negative and bypassing the check for a large value.
Without this check, a remote attacker can make a crafted file with a large \texttt{item\_num} field such as \texttt{0xffffffff}, causing %the integer variable to overflow and turn into a negative value, triggering 
a buffer overflow issue
%causing a buffer overflow by function \texttt{avio\_read} on Line 28. 
%Through an assignment in Line 27, this manipulation triggers a buffer overflow 
in the \texttt{avio\_read} function at Line 25 and potentially leads the application to exhibit incorrect behavior or crash. %\todo{say the vulnerable statements line number}

V-SZZ~\mbox{\cite{bao2022v}} assumes the deletion lines in the security patch as vulnerable codes and the basic idea of it is to pinpoint the earliest commit introducing the deletion lines as \mbox{\textit{vic}}. However, vulnerability logic is composed of various vulnerable codes, such as Lines 21 and 25 in Listing~\mbox{\ref{list:CVE-2017-14169}}. V-SZZ only considers tracing back deleted line (Line 12) and overlooks the real vulnerability behavior, leading to incorrect identification of \mbox{\textit{vic}}.
V0Finder~\mbox{\cite{woo2021v0finder}} attempts to identify vulnerable versions using code clone detection. It generates hash values for the entire vulnerability function and identifies code clones according to the distance value of two hashes. This method may still introduce excessive vulnerability-unrelated features due to the inclusion of ``entire function''.

\begin{lstlisting}[
	style = interfaces,
    % style=nonbreak,
    caption = {Motivating Example of CVE-2017-14169.},
    label = {list:CVE-2017-14169}]
diff --git a/libavformat/mxfdec.c b/libavformat/mxfdec.c
@@ -493,11 +493,11 @@ static int mxf\_read\_primer\_pack 
static int mxf_read_primer_pack(void *arg, AVIOContext *pb, int tag, int size, UID uid, int64_t klv_offset)
{
    MXFContext *mxf = arg;
    int item_num = avio_rb32(pb);
    int item_len = avio_rb32(pb);
    if (item_len != 18) {
        avpriv_request_sample(pb, "Primer pack item length %d", item_len);
        return AVERROR_PATCHWELCOME;
    }
`\colorbox{brightmaroon!30}{-\ \ if (item\_num > 65536) \{}`
`\colorbox{grannysmithapple!50}{+\ \ if (item\_num > 65536 || item\_num < 0) \{}`
        av_log(mxf->fc, AV_LOG_ERROR, "item_num %d is too large\n", item_num);
        return AVERROR_INVALIDDATA;
    }
    if (mxf->local_tags)
        av_log(mxf->fc, AV_LOG_VERBOSE, "Multiple primer packs\n");
    av_free(mxf->local_tags);
    mxf->local_tags_count = 0;
    mxf->local_tags = av_calloc(item_num, item_len);
    if (!mxf->local_tags)
        return AVERROR(ENOMEM);
    mxf->local_tags_count = item_num;
    avio_read(pb, mxf->local_tags, item_num*item_len);
    return 0;
\end{lstlisting}

% \iffalse
% %https://github.com/ffmpeg/FFmpeg/commit/757da974b21833529cc41bdcc9684c29660cdfa8
% \begin{lstlisting}[
%     style=interfaces,
%     caption = {Motivating Example of Solely adding Commit.},
%     label = {list:addition}, aboveskip=-3pt, belowskip=-3pt]
% diff --git a/libavcodec/g729_parser.c b/libavcodec/g729_parser.c
% index 8c06ce4ee6aec..4dcdeab651c6b 100644
% --- a/libavcodec/g729_parser.c
% +++ b/libavcodec/g729_parser.c
% @@ -48,6 +48,9 @@ static int g729_parse(AVCodecParserContext *s1, AVCodecContext *avctx,
%   if (avctx->codec_id == AV_CODEC_ID_ACELP_KELVIN)
%       s->block_size++;
% `\colorbox{grannysmithapple!50}{+\ if (avctx->channels > 2)}`
% `\colorbox{grannysmithapple!50}{+\ \ \ \ s->block\_size = 0;}`
%   s->block_size *= avctx->channels;
%   s->duration = avctx->frame_size;
% }
% \end{lstlisting}
% \fi

\vspace{4px}
\noindent\textbf{Observation.} Identification of vulnerable versions is hindered by the challenge of comprehending vulnerability logic from the security patch. LLMs can grasp contextual code semantics and achieve human-like understanding in code-related tasks, eliminating the reliance on predefined patterns and rules utilized by traditional static analysis and automated tools. Through the integration with an LLM, we can automate vulnerability analysis, enhancing the efficiency of our approach.
%Consequently, \textsc{Vercation} employs vulnerability-fixing commit analysis to extract crucial vulnerable statements that accurately portray the logical behavior of the vulnerability. Additionally, it incorporates customized weight allocation strategies to mitigate the impact of statements unrelated to the vulnerability, thereby minimizing noise.
%This underscores the imperative to unveil vulnerability behavior directly from the patch. Consequently, \textsc{Vercation} employs vulnerability-fixing commit analysis to extract pivotal vulnerable statements that accurately portray the logical behavior of the vulnerability. Additionally, it incorporates customized weight allocation strategies to mitigate the impact of statements unrelated to the vulnerability, thereby minimizing noise.

%https://github.com/ffmpeg/FFmpeg/commit/b7702fafb35
\begin{lstlisting}[
    style=interfaces,
    caption = {Motivating Example of Code Refactoring Commit.},
    label = {semantic}, aboveskip=-3pt, belowskip=-3pt]
diff --git a/libavformat/avidec.c b/libavformat/avidec.c
@@ -350,8 +350,7 @@ static void avi_read_nikon(AVFormatContext *s, uint64_t end)
  uint16_t tag     = avio_rl16(s->pb);
  uint16_t size    = avio_rl16(s->pb);
  const char *name = NULL;
  char buffer[64]  = { 0 };
`\colorbox{brightmaroon!30}{-\ if (avio\_tell(s->pb) + size > tag\_end)}`
`\colorbox{brightmaroon!30}{-\ \ \ \ size = tag\_end - avio\_tell(s->pb);}`
`\colorbox{grannysmithapple!50}{+\ size = FFMIN(size, tag\_end - avio\_tell}` 
`\colorbox{grannysmithapple!50}{+\ \ \ \ (s->pb));}`
  size -= avio_read(s->pb, buffer, FFMIN(size, 
      sizeof(buffer) - 1));

/*The definition of method FFMIN in libavutil/common.h*/
#define FFMIN(a, b) ((a)>(b)?(b):(a))
\end{lstlisting}

%补丁通过
\textbf{Example 2)} 
%We noticed many modifications of refactoring commits where the code has the same semantics but with low syntactic similarity between the pre-modification and post-modification commits.
After extracting the vulnerability logic and its corresponding vulnerable statements ($S_v$), we can locate the \textit{vic} by backtracing the code changes of $S_v$ in the previous commit, thereby identifying the vulnerable versions.  
%During the backtrack from \textit{pc} to \textit{vic}, certain commits may modify the code we are tracking. 
V-SZZ~\cite{bao2022v} identifies code clones before and after such code changes by the edit distance. However, refactoring commits, prevalent in OSS development, aim to optimize and reorganize code while preserving its functionality~\cite{du2004refactoring,almogahed2023refactoring}. %Such commits can result in situations where the code's semantics remain the same, but there is a significant syntactic difference between the code before and after modifications. Essentially, such a commit would cause the tool like V-SZZ to incorrectly flag it out as a \textit{vic}. %even though the actual vulnerability may have been introduced in another previous commit. }
%This poses challenges for approaches using the Levenshtein distance algorithm, which is primarily designed to handle syntactic differences.}

A refactoring commit in FFmpeg (Listing~\ref{semantic}) demonstrates how structural changes caused by method encapsulation mislead existing detection methods. In this commit, developers extracted the code with \texttt{if} condition structure into a new method \texttt{FFMIN()} to improve code modularity. While this refactoring preserves functionality, it introduces significant syntactic divergence between the original code (Lines 7-8) and the restructured code (Lines 9-10).

Traditional tools like V-SZZ~\cite{bao2022v} consider edit distance as the metric to detect code clones. The edit distance between the original and refactored code drops to 48\% due to the method outlining, causing V-SZZ to incorrectly flag this commit as the vulnerability-introducing commit (\textit{vic}). Other approaches like ReDeBug~\cite{jang2012redebug} and V0Finder~\cite{woo2021v0finder} also fail to recognize the equivalence between inline logic and encapsulated methods, as their coarse pattern matching ignores structural abstraction.

\vspace{4px}
\noindent\textbf{Observation.} This case highlights the structural refactoring in OSS development and the limitations of existing clone detection techniques. \textsc{Vercation} addresses this by expanding function calls during AST generation—inlining the \texttt{FFMIN()} method body to reconstruct the original logic. Combined with AST normalization, this allows \textsc{Vercation} to detect code logic equivalence despite structural variations, accurately tracing the \textit{vic} through refactored commits.

\section{Design of \textsc{Vercation}}
\begin{figure*}[t]
    \centering
	\includegraphics[width=0.75\linewidth]{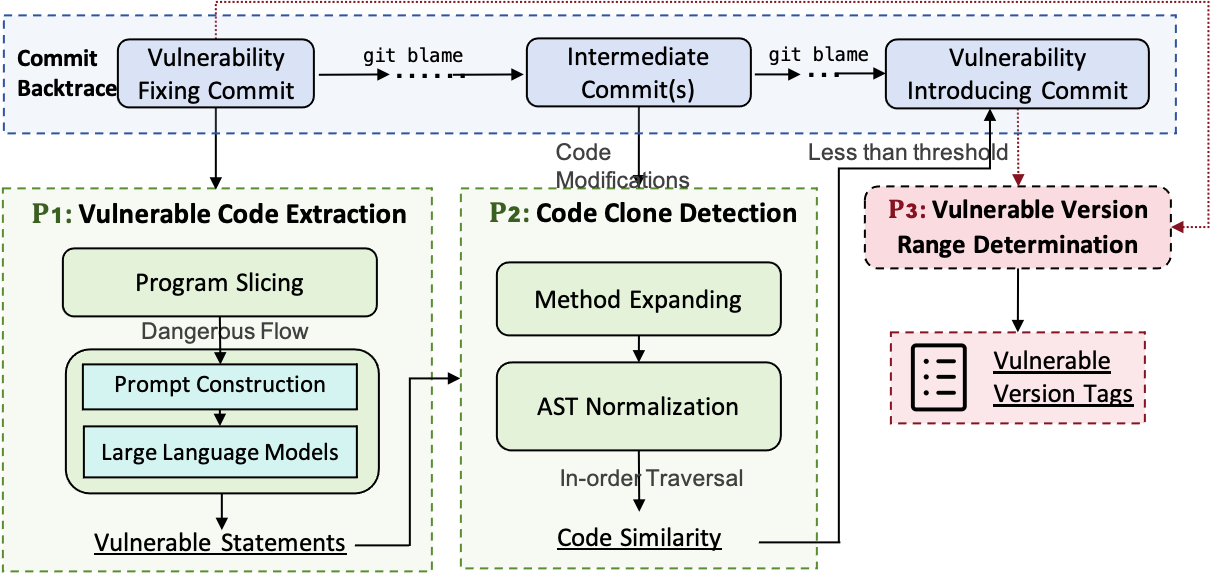}  
  	\caption{\textsc{Vercation} Workflow.}
  	\label{fig:workflow}
\end{figure*}

We propose \textsc{Vercation}, an end-to-end automated approach designed for vulnerable version identification in OSS vulnerabilities.
The high-level workflow is illustrated in Figure~\ref{fig:workflow}. \textsc{Vercation} consists of three phases: \textit{Vulnerable code extraction} (P1), \textit{Code clone detection} (P2) and \textit{Vulnerable version range determination} (P3).

In P1, \mbox{\textsc{Vercation}} combines program slicing and LLM to identify and extract vulnerability-related program statements in a precise manner. Specifically, we utilize the patch code as a slicing criterion to extract \textit{dangerous flows}, defined as program statements that directly or indirectly affect the variables or expressions at the patch code. To mitigate the risk of including statements unrelated to the vulnerability (false positives), we employ prompt engineering with few-shot and Chain-of-Thought (CoT) strategies, empowering LLM to reason with the vulnerability based on the extracted dangerous flows and accurately extract vulnerability-related statements.

In P2, \textsc{Vercation} retraces historical commits to collect previous modifications of vulnerable statements. For each statement before and after modification, \textsc{Vercation} expands the functions within the statement, generates ASTs, and normalizes them. Then we utilize an in-order traversal algorithm to compare the AST before and after the modification as code similarity, which determines if the commit is the initial introduction of the vulnerable statements.

In P3, \textsc{Vercation} identifies the affected versions based on the CVE's \textit{vfc} and the corresponding \textit{vic}.

%In the following sections, we explain these phases in detail. 
\hl{To note, \textsc{Vercation} supports patches that span multiple functions and files. In P1, we extract dangerous flows from each affected function independently, then combine all extracted dangerous flows as input to the LLM for unified vulnerability logic analysis. In P2, we perform a backtrack analysis for vulnerable statements from all affected functions.}

\subsection{Vulnerable Code Extraction}
\label{sec:vulextract}
Taking a \textit{vfc} as an input, \textsc{Vercation} identifies and extracts program statements related to a vulnerability in three steps, as explained in the following subsections respectively.
%We generate vulnerability logic from vulnerability-fixing commits powered by LLMs and extract vulnerable lines as signatures in three steps as explained in the next three subsections respectively.

\subsubsection{Dangerous Flow Extraction}
Firstly, \textsc{Vercation} conducts program slicing~\cite{weiser1984program} on the source code to extract crucial program statements that may contribute to the vulnerability, which we call \textit{dangerous flow}. The slicing is based on the following criteria: deleted/added statements and patch-related variables. 
%We define the context that has dependencies on patch-related variables as crucial vulnerable statements \todo{$S_c$}. 
Building upon existing techniques~\cite{xiao2020mvp, shi2022precise, woo2022movery}, we execute program slicing on the program dependency graph of the patched function according to the slicing criteria. 
\hl{We performed slicing in two directions: backward and forward.
\textit{Backward slicing} is used to trace the source of patch-related variables and \textit{forward slicing} is used to find the trigger behavior that causes the vulnerability.}
For different statement types, we customize different slicing rules:
\begin{itemize}[leftmargin=*]
\item \textit{Assignment statement} affects the data-flow values of vulnerability behavior. We conduct normal slicing in the assignment statements and the assigned variables should be added to the slicing criterion. For example, if we take \texttt{item\_num} of Line 12 and 13 in Listing~\ref{list:CVE-2017-14169} as slicing criterion and perform forward slicing, Line 21 and 24 are included.
	\item \textit{Conditional statement} affects the reaching condition of the vulnerability trigger. We aim to slice the condition statement that takes the variables of slicing criterion as condition check, and the result also includes all the subsequent statements of condition statement (e.g., subsequent statement Line 14-15 of condition statement Line 13, subsequent statement Line 23 of condition statement Line 22).
	\item \textit{Function call statement.} If the function call statement's parameters contain patch-related variables included in the slicing criterion, we conduct slicing on the statement.
	\item \textit{Return statement.} There is no need for forward slicing because there is no dependency between the return value and the statements following the return statement. For example, there is no need for forward slicing on Line 15 and 23 in Listing~\ref{list:CVE-2017-14169}.
\end{itemize}

%Meanwhile, the slicing is performed in two directions: backward and forward.
%\textit{Backward slicing} is to trace the source of patch-related variables. For example, when we designate \texttt{item\_num} (e.g., Line 12, 13 in Listing~\ref{list:CVE-2017-14169}) as the criterion, the result of the backward slice is Line 6 (Line 12 is data-dependent on Line 6). Additionally, Line 12 is control-dependent on Line 8, as its execution is contingent on the unsatisfied condition in Line 8; otherwise, the function will promptly return. Following backward slicing, we retain Lines 6 and 8 as backward dangerous flow.
%\textit{Forward slicing} is used to find the trigger behavior that causes the vulnerability. For example, in Listing~\ref{list:CVE-2017-14169}, if we set the modified statements Lines 12 and 13 along with the variable \texttt{item\_num} as criteria, the outcomes of forward slicing include 1) Lines 21, 24 and 25, which are directly or indirectly data-dependent on the variable \texttt{item\_num} and 2) Lines 14, 15, 22 and 23, which are directly or indirectly control-dependent on the variable \texttt{item\_num}. 

\subsubsection{Dangerous Flow-based Prompt Construction}
Secondly, we leverage an LLM to refine these dangerous flows further since the dangerous flows extracted in the above step (program slicing) may include false positives (unrelated to vulnerability logic). 
We explore prompting strategies to assist LLM in the vulnerability comprehension task and in refining vulnerable statements from the dangerous flows extracted through program slicing.
\hl{It is important to note that the LLM's role is to perform code understanding and vulnerability logic analysis rather than to recall specific CVE information from its training data. Most vulnerability reports do not provide detailed vulnerability logic analysis or identify specific vulnerable code statements}~\cite{aghaei2023automated}, \hl{where the LLM's semantic analysis capabilities become essential.}

Our method utilizes an LLM that supports interaction with system prompts and user prompts. The system prompt sets the role and background of the LLM. The user prompt consists of specific instructions issued by the user.
At the beginning of the system prompt, we assign the role of a security researcher to the model with the statement, ``You are a security researcher an expert in detecting security vulnerabilities," and indicate that we will provide the CVE information and dangerous flow. The prompt also concludes with a declaration of the fixed output format expected for the model's response.
The content of the user prompt includes detailed CVE information including CVE ID, CWE ID, CVE description, and dangerous flow extracted from program slicing.
\hl{The extracted dangerous flows are formatted with line numbers prefixed to each statement when constructing prompts, such as ``\texttt{4442 total\_size += msec->size; 4444 stash->info\_ptr\_memory = (bfd\_byte *) bfd\_malloc (total\_size);}''.}

\begin{figure}
    \centering
    \includegraphics[width=0.98\linewidth]{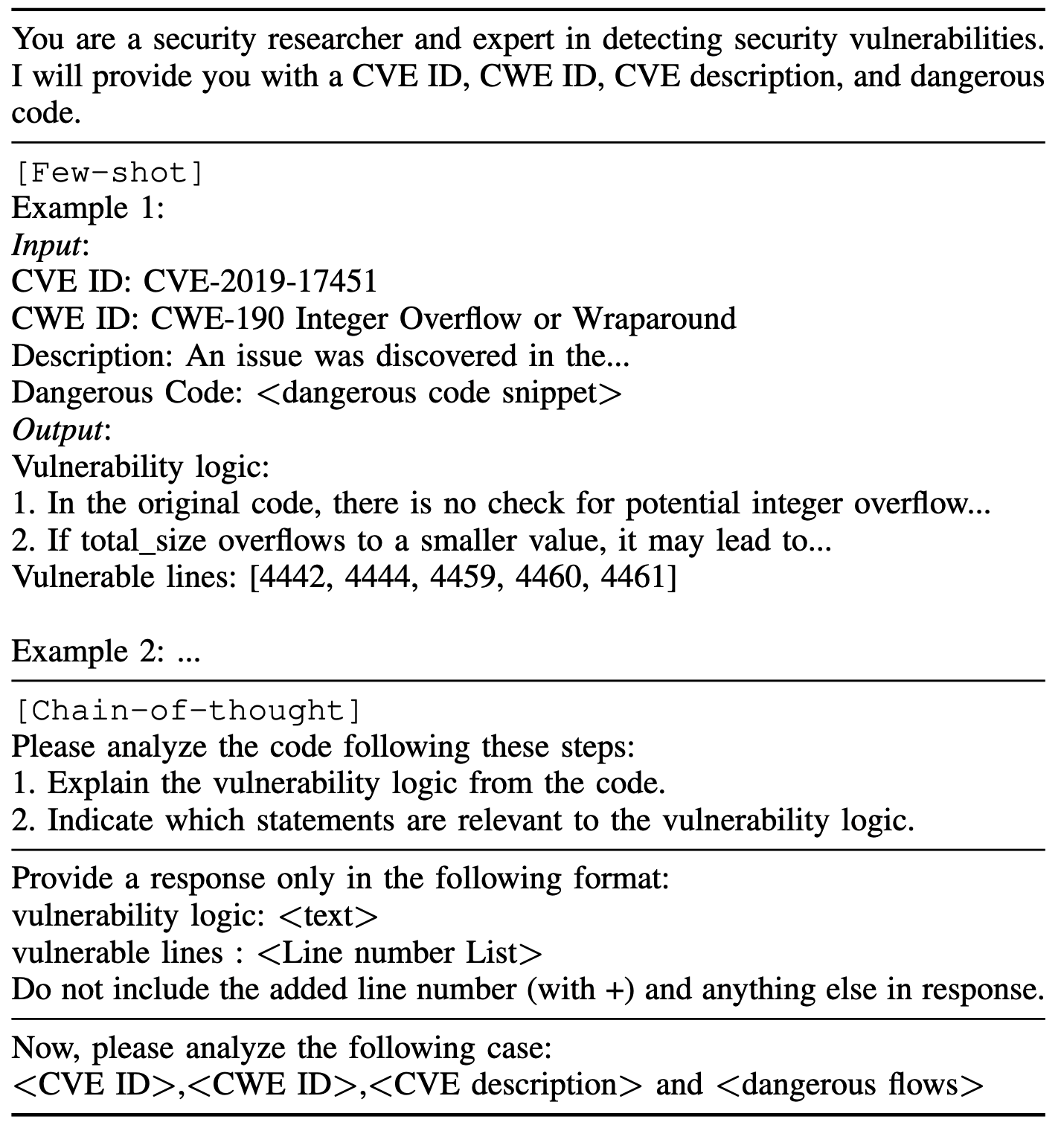}
    \caption{The LLM Prompt Structure.}
    \label{fig:prompt}
\end{figure}

Additionally, we use the following prompt strategies to optimize the performance of LLM:
\begin{enumerate}[label=(\roman*),leftmargin=*]
\item{ 
%Few-shot prompting is a technique where we provide the LLM with a small number of examples demonstrating the desired task before presenting the actual problem. This approach helps guide the model's behavior by showing it how to respond to similar inputs.
%We include two carefully crafted examples in the user prompt before providing the target CVE information.
%Each example consists of:
%a) A sample CVE ID, CWE ID and CVE description. 
%b) Corresponding dangerous flow extracted from %program slicing.
%c) The expected vulnerability logic analysis.
%d) The correctly identified vulnerable statements.
%To avoid data leakage, the examples are not contained in our experiment dataset.

Few-shot prompting is a technique where we provide the LLM with a small number of examples demonstrating the desired task before presenting the actual problem. 
We carefully select two examples from publicly available and verified CVE cases. To avoid data leakage, the cases are not included in our experimental dataset. The selection of examples follows specific criteria: (1) having complete and clear CVE descriptions, (2) containing explicit vulnerability fix code, and (3) featuring easily understandable vulnerability logic. These examples were created and reviewed through a systematic process: analyzing CVE descriptions and patch code, extracting key vulnerability logic, annotating relevant vulnerable statements, and writing clear vulnerability analysis explanations. The examples were created by the first author and reviewed by the fourth and fifth authors. To ensure experimental reproducibility, we maintain a fixed set of examples throughout all experiments.
Each example consists of: a) A sample CVE ID, CWE ID and CVE description, b) Corresponding dangerous flow extracted from program slicing, c) The expected vulnerability logic analysis, and d) The correctly identified vulnerable statements.
}
\item{Chain-of-thought (CoT) prompting provides LLM with a prompt that encourages it to generate intermediate reasoning steps before arriving at a final answer~\cite{wei2022chain}. 
We prompt the model to ``Please analyze the code following these steps:
1. Explain the vulnerability logic from the code
2. Indicate which statements are relevant to the vulnerability logic". %and require the response in a specific two-part format:
%vulnerability logic and vulnerable lines.
This structured format naturally implements a chain of thought - requiring the model to first reason about and explain the vulnerability logic (the reasoning step) before identifying the specific vulnerable lines (the identification step). 
\hl{This approach identifies the vulnerable statements using the code understanding capability of LLM rather than superficial pattern matching.}
%This approach ensures that the model's identification of vulnerable statements is based on a thorough understanding of the vulnerability pattern rather than superficial pattern matching.
Table~\ref{fig:prompt} shows our prompting template, and we take the vulnerable statements from the response as the input for the next step.}
\end{enumerate}

\iffalse
\begin{table}[t]
\centering
\caption{The LLM Prompt Structure.}
\begin{tabular}{@{}p{1\columnwidth}@{}}
\toprule
    You are a security researcher and expert in detecting security vulnerabilities.
    \\
    I will provide you with a CVE ID, CWE ID, CVE description, and dangerous code.
    \\
\midrule
\texttt{[Few-shot]}
\\
Example 1:
\\
\textit{Input}:
\\
CVE ID: CVE-2019-17451
\\
CWE ID: CWE-190 Integer Overflow or Wraparound
\\
Description: An issue was discovered in the...
\\
Dangerous Code:
$<$dangerous code snippet$>$
\\
\textit{Output}:
\\
Vulnerability logic:
\\
1. In the original code, there is no check for potential integer overflow...
\\
2. If total\_size overflows to a smaller value, it may lead to...
\\
Vulnerable lines: [4442, 4444, 4459, 4460, 4461]

\\
Example 2: ...\\
\midrule
    \texttt{[Chain-of-thought]}
    \\
   Please analyze the code following these steps:
   \\
   1. Explain the vulnerability logic from the code.
   \\
   2. Indicate which statements are relevant to the vulnerability logic.
    \\
\midrule
    Provide a response only in the following format: 
    \\
    vulnerability logic: $<$text$>$
    \\
    vulnerable lines : $<$Line number List$>$
    \\
    Do not include the added line number (with +) and anything else in response.
    \\
    \midrule
Now, please analyze the following case: 
\\
$<$CVE ID$>$,$<$CWE ID$>$,$<$CVE description$>$ and $<$dangerous flows$>$
\\
\bottomrule

\end{tabular}
\label{tab:prompt_1}
\end{table}

\fi

\subsection{Code Clone Detection}
\label{sec:codechange}

\begin{figure*}[t]
    \centering
    \includegraphics[width=0.7\linewidth]{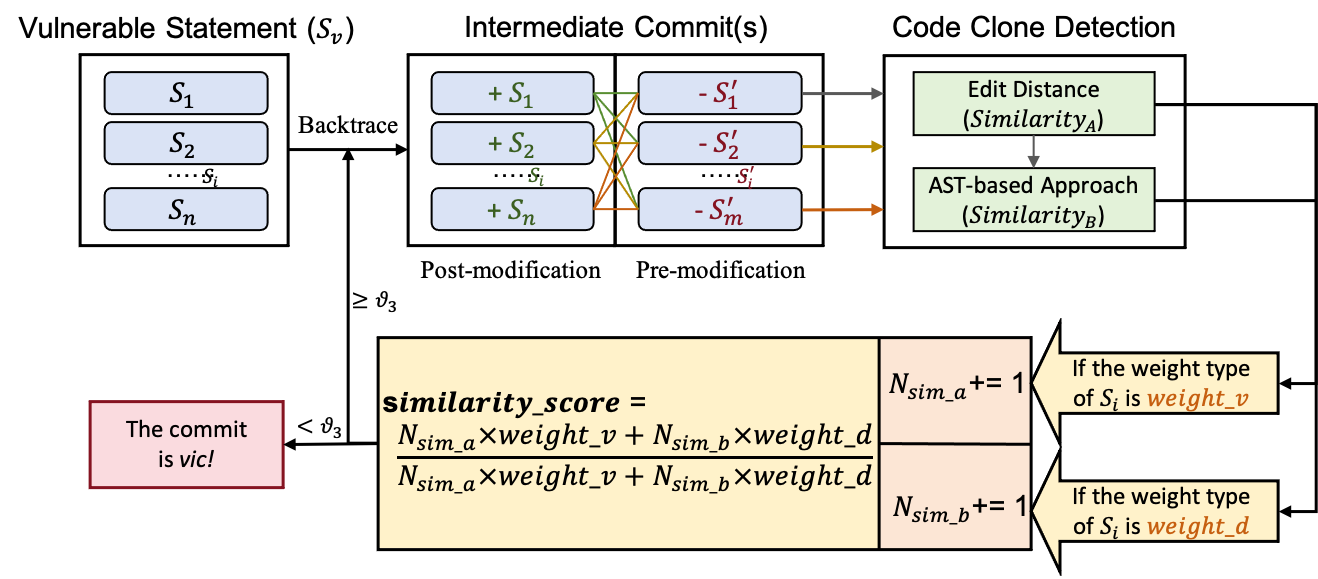}
    \caption{Overview of Code Clone Detection Process in \textsc{Vercation}.}
    \label{fig:codeclone}
\end{figure*}

Given the vulnerable statements $S_v$ of a fixing commit (extracted in Section~\ref{sec:vulextract}), \textsc{Vercation} performs a commit backtrack to identify which previous commit introduced $S_v$ by tracing the history of earlier modifications.
To achieve this, \textsc{Vercation} employs the \texttt{git blame} command to backtrack through previous commits.
Figure~\ref{fig:codeclone} illustrates our code clone detection process. For each intermediate commit between the \textit{vfc} and the potential \textit{vic}, we compare the post-modification statements (\var{S}) with the pre-modification statements (\var{S'}) using two levels of similarity detection: a quick syntactic check using edit distance and a deeper structural analysis using ASTs.
The results from both comparisons, combined with statement weights assigned in Section~\ref{sec:statementweight}, contribute to a final similarity score. 
If this score is below the threshold \var{\theta_3}, we identify the commit as the \textit{vic}. Otherwise, we continue the backtrack process.

\subsubsection{Syntactic Similarity Analysis}
First, we use line mapping to pre-filter highly similar lines in \var{S'} with \var{S}, where $S \subseteq S_v$. 
If such lines are found, it indicates that the commit is not the first to introduce \var{S_v}.
We use the edit distance to calculate line similarity $\mathrm{Similarity}_{A}(S'_i, S_i)$, where $S'_i \in S'$ and $S_i \in S$. 
We set a similarity threshold $\vartheta_1$; if $\mathrm{Similarity}_{A}(S'_i, S_i) \ge \vartheta_1$, we continue to backtrack the previous commit. 
If $\mathrm{Similarity}_{A}(S'_i, S_i) < \vartheta_1$, we proceed to the more detailed structural analysis.
\hl{To note, when encountering merge commits with multiple parents during backtrack, we traverse each parent path independently and select the chronologically earliest \textit{vic} across all branches.}

% \noindent\textbf{AST Expansion and Normalization.}
%要做语义分析的原因 example
\subsubsection{Structural Similarity Analysis}
As exemplified in Example 2 of Section~\ref{sec:motivating}, Lines 5-6 (deleted pre-commit lines) and 7-8 (added line of post-commit, e.g., included in $S_v$) exhibit high similarity in code behavior. However, previous research failed to identify this high similarity through simple line mapping, resulting in the commit (version) being erroneously labeled as an introducing commit (vulnerable version). % statements to $S_v$ in the pre-modification commit by simple syntactic analysis, mistakenly identifying the commit as introducing commit.
We propose a fine-grained similarity comparison method at the AST-level to address this limitation. AST is a tree representation of code that preserves well-defined components of statements, explicit statement order and the execution logic~\cite{yang2021asteria}. 

\noindent\textbf{Method Expansion.} Due to code refactoring often outlining code into a new method, as shown in Listing~\ref{semantic}, the codes in Lines 7-8 are outlined into methods \texttt{FFMIN}. \hl{The standard AST comparison treats the if-condition (Lines 7-8) as completely different from the FFMIN function call, resulting in a false vulnerability identification. Therefore, we utilize the inline technique to expand the method during AST generation to gather more code behavior within the statements.} This involves obtaining the method definition at the callsite and expanding the method body inline, resulting in a more comprehensive code representation. 
We get the AST for the target file using \textit{Clang} without performing actual compilation.
%Additionally, we perform syntax checking without actual compilation. Combining the collected dependency information in step 1) as the compiler option of \textit{Clang}, we can get the  AST for the target file. 
Furthermore, through recursive retrieval of line numbers for sub-ASTs, we can locate the sub-AST corresponding to $S_v$. %通过AST 中的line number定位到Sc的sub AST

\vspace{4px}
\noindent\textbf{AST Normalization.}
%Since the code programming style is variable, separate codes may perform semantic similarity. 
To address structural variations arising from different developer coding styles, it is essential to normalize the AST. 
This process mitigates inconsistencies and reduces analytical noise by standardizing code representations.
AST normalization involves conditional structure normalization, loop structure normalization, and relational operation normalization.  

% 并进行多个阶段的标准化工作，其中包括选择语句标准化、循环语句标准化、函数调用结构标准化等，最终得到本文分析使用的语法树；
\begin{itemize}[leftmargin=*]
	\item \textit{Conditional structure} can be implemented by \textit{if-else} and \textit{switch-case}. We normalize the structure of \textit{switch-case}: \texttt{switch(expression)\{case value1: statement1; break; case value2: statement2; break; default statement3;\}} into a \textit{if-else} structure: \texttt{if(exp == value1) statement1; else if(exp == value2) statement2; else statement3}.
	\item \textit{Loop structure} can be represented as \textit{while}, \textit{do-while} and \textit{for} structures, we normalize all the loop structure to \textit{while}~\cite{xue2022clone}. The \textit{for} structure is \texttt{for(initialize; condition; increment) \{statements;\}}, we transform it into \textit{while} structure: \texttt{initialize; while(condition) \{statements; increment\}}.
	\item \textit{Relational operation} include commutative and noncommutative operation. Commutative relational operation (i.e., +, !=, \&\&) refers to the operation where commutate the operands does not change the operation semantics. We normalize the two operands (left and right child nodes) 
    of the operation (parent nodes) by alphabetical order. For example, we transform $b + a$ (b is the left child node) to $a + b$ (a is the left child node). Noncommutative relational operation (i.e., $>$, $<=$) will change the operation semantic when commutating the operands, we normalize them by predefined rules. For example, we transform $b < a$ to $a > b$ (normalize operation $<$ to $>$).
\end{itemize}

\vspace{4px}
\noindent\textbf{AST Similarity.} 
Ultimately, we utilize an in-order traversal to ascertain whether the commit introduces \var{S} by comparing the AST similarity of \var{S_i} and \var{S'_i}. First, we perform an in-order traversal of the ASTs for \var{S_i} and \var{S'_i}, converting the results into sequences \var{Sq_i} and \var{S'q_i}. Then, we calculate the edit distance between the two sequences as their AST-based similarity $\mathrm{Similarity}_{B}(Sq_i, S'q_i)$. 
We set a similarity threshold $\vartheta_2$, and if $\mathrm{Similarity}_{B}(Sq_i, S'q_i) \ge \vartheta_2$, $S_i$ and $S'_i$ are considered to be AST-based similar. 
If $\mathrm{Similarity}_{B}(Sq_i, S'q_i) < \vartheta_2$, it indicates that there is no statement similar to $S_i$ in the pre-modification commit.
During this process, we classify a commit as a refactoring commit when it exhibits low edit-distance similarity ($\mathrm{Similarity}_{A}(S_i, S'_i) < \vartheta_1$) but high AST-based similarity ($\mathrm{Similarity}_{B}(Sq_i, S'q_i) \ge \vartheta_2$). This indicates that while the code structure has been significantly modified, the underlying functionality remains unchanged.

\subsubsection{Statement Weight Allocation}
\label{sec:statementweight}
%statement权重分配原则
% As vulnerable statements contribute differently to vulnerability triggers, we propose strategies to allocate weights to the vulnerable statements extracted by LLM. 
Not all statements identified by an LLM contribute equally to vulnerability in triggering the exploit. 
To address this, we propose a weight allocation strategy that prioritizes vulnerable statements based on their invocation of known sensitive functions. 
The core premise is that the presence of specific, high-risk functions within a statement is a strong indicator of its criticality.
% We have noticed that certain types of C/C++ vulnerabilities often involve specific functions. 
%\sethlcolor{light-gray}\hl{Through a systematic process involving three authors with security expertise and incorporating findings from relevant papers}~\cite{table1,table2}, \sethlcolor{light-gray}\hl{we created a pre-defined table (Table}~\ref{tab:dangerous_function}) \sethlcolor{light-gray}\hl{of sensitive functions for various vulnerability types. The classification was initially drafted and then validated by authors with extensive experience in vulnerability analysis. For instance, in Listing 1, Lines 24 and 28 are critical trigger lines. With this knowledge, we can rely on Table 1 to identify such functions within vulnerable statements extracted by LLM.}
We followed a systematic process to identify and categorize these sensitive functions. Starting with a comprehensive review of previous studies~\cite{table1,table2}, we collected existing sensitive function enumerations. We then conducted a preliminary study of common C/C\texttt{++} vulnerabilities to identify the most frequent vulnerability categories, such as buffer overflow, integer overflow, and use-after-free, and examined the functions commonly involved in known vulnerabilities for each type. This process involved three authors with security expertise: the first author proposed the initial categorization, which was then independently reviewed and validated by the fourth and fifth authors, with any discrepancies resolved through discussion. 
The result is the pre-defined table (Table~\ref{tab:dangerous_function}) of sensitive functions for various vulnerability types. 
We analyze each vulnerable statement extracted by the LLM. 
If a statement contains a direct call to any function listed in our sensitive function table, it is assigned a higher weight, marking it as a probable vulnerability trigger.
For instance, in Listing~\ref{list:CVE-2017-14169}, Lines 21 and 25 are critical trigger lines. 
% With this knowledge, we can rely on Table~\ref{tab:dangerous_function} to identify such functions within vulnerable statements extracted by LLM.

%For instance, in Listing 1, Lines 24 and 28 are critical trigger lines. 
%With this knowledge, we can rely on a pre-defined table (Table~\ref{tab:dangerous_function}) to identify such functions within vulnerable statements extracted by LLM. 
%We summarized the common types of vulnerabilities in C/C++, and combined the authors' domain knowledge with the enumeration of sensitive functions from relevant papers~\cite{table1,table2} to create the pre-defined table.
% Some OSS commonly used customized functions serve as variants of sensitive functions.  

Furthermore, we conduct an inter-procedural analysis to identify customized functions that serve as variants of canonical sensitive functions.
For instance, the \texttt{av\_calloc} function on Line \hl{21} is a variant of the \texttt{calloc} function. 
To discern whether the customized function \texttt{F} contains sensitive functions, 
we first analyze the definition and implementation code of function \texttt{F}, then check if \texttt{F}'s function body calls any sensitive functions from Table~\ref{tab:dangerous_function}. 
If such call relationships are found, \texttt{F} is also identified as a sensitive function.
Taking Listing~\ref{list:CVE-2017-14169} as an example, considering this is an overflow-type vulnerability, we recognize the \texttt{av\_calloc} in Line \hl{21} and \texttt{avio\_read} function in line \hl{25} as sensitive functions.

%In the preceding step, the extracted vulnerable statements may be unrelated to vulnerability trigger behavior. For example, considering Line 17-18, 26 and 27 in Listing~\ref{list:CVE-2017-14169} as crucial vulnerable statements would introduce excessive noise, as these statements are not highly relevant to the vulnerability trigger behavior. To mitigate this issue, we propose a statement-level weight allocation strategy aimed at minimizing noise.
% Algorithm 1 shows how to weight statement-level code by using code inspection approach. Initially, the algorithm assigns a default weight to all statements. Then we apply two allocation strategies to weight the statements.

\begin{table}[tp]
	\caption{Sensitive Functions for Some Vulnerability Types.}
	\begin{center}
		\begin{threeparttable}
			\begin{tabular}{l l}
				\toprule
				\textbf{Vulnerability Type} & \textbf{Sensitive Function} \\
				\midrule
				Buffer overflow & \makecell[l]{strcpy, strncpy, memcpy, memset, read, \\ write, gets, gets\_s, strcat} \\
                \midrule
				Integer overflow & add, multiple, bit-shifting, memcpy \\
                \midrule
				Null pointer derefrence & malloc, calloc, realloc, strdup \\
                \midrule
				Pointer out-of-bounds access & memcpy, memmove \\
                \midrule
				Use after free & free, malloc, calloc\\
                \midrule
				Format string & sprintf, printf, scanf \\
                \midrule
				Arbitrary command execution & fopen, popen, system \\
				\bottomrule
			\end{tabular}
		\end{threeparttable}
	\end{center}
	\label{tab:dangerous_function}
\end{table}

\vspace{4px}
\noindent\textbf{Parameters.} There are two parameters for the weight allocation of statements, i.e., $W_{sensitive}$ and $W_{base}$, that could affect the effectiveness of \textsc{Vercation}: 
1) $W_{base}$ is the base weight for initializing the weights of all statements.
2) $W_{sensitive}$ is the weight for the statements containing sensitive functions.
We set $W_{sensitive} = 2 \times W_{base}$, $W_{base}=1.0$, the detailed weighting configuration experiment is in Section~\ref{sec:evaluation_llm}. %Ultimately, we consider the statements in $S_c$ with a weight greater than \textit{weight\_base} as vulnerability signatures $S_v$.

% 
% \vspace{4px}
% \noindent\textbf{Code Clone Determination.}
%and we proceed the commit backtracking process. If $Similarity_B(Sq_c, Sq_i^{'}) < \vartheta_2$, it means that there is no statement similar to $S_c$ in the pre-modification commit, so the commit introduced $S_c$, which is \textit{vic}. 
\subsubsection{Code Clone Determination}
We calculate the \textit{similarity\_score} for each code modification of the previous commit with the backtrack. 
We set a threshold $\vartheta_3$, if $similarity\_score < \vartheta_3$, we determine that there is no code clone between the pre-modification code of this previous commit and the vulnerable statements, thus confirming that this commit introduced the vulnerability.
Otherwise, we proceed with the commit backtrack process.
%$$Totalweight = a\times weight\_v + b \times weight\_d$$

\begin{equation}
\begin{split}
similarity\_score = \frac{sim\_a\times W_{sensitive} + sim\_b \times W_{base}}{a\times W_{sensitive} + b \times W_{base}} \nonumber
\end{split}
\end{equation}
where, 
$W_{sensitive}$ and $W_{base}$ are predefined weights assigned to different types of statements;
\var{a} and \var{b} are the total counts of the statements in \var{S_v} corresponding to these weight types;
% \texttt{sim\_statement} refers to the statements where $\mathrm{Similarity}_{A}(S_i, S'_i)) \ge \vartheta_1$ or $\mathrm{Similarity}_{B}(Sq_i, S'q_i) \ge \vartheta_2$ (i.e., edit-distance or AST-based similarity) during each commit comparison. 
$sim\_a$ and $sim\_b$ are the counts of statements that were found to be similar, either through syntactic or structural analysis.

\subsection{Vulnerable Version Range Determination}
To precisely determine the range of software versions affected by a vulnerability, we employ a methodology based on the principle of commit reachability.
The core premise is that a vulnerability, once introduced in a specific \textit{vic}, is propagated to all subsequent commits in its lineage until it is fixed by a \textit{vfc}.

The basis for our analysis is the concept of reachability, where a commit \texttt{A} is considered reachable from another commit \texttt{B}, if commit \texttt{A} is an ancestor of commit \texttt{B} in the project's commit graph~\cite{bao2022v}. 
As version tags are essentially pointers to specific commits, they inherit the reachability properties of the commits they represent.
This allows us to map the vulnerability's presence from the commit level to software versions.

Our methodology formally defines the set of vulnerable versions ($V_v$) as follows: First, we identify $V_i$, the set of all version tags reachable from \textit{vic}. Second, we identify $V_f$, the set of all version tags reachable from \textit{vfc}. The vulnerable version set $V_v$ is then derived from the set difference between these two sets: $V_v = V_i - V_f$.

We illustrate the procedure by the case of \texttt{CVE-2021-20294} in the Binutils project~\cite{CVE-2021-20294}.
For this vulnerability, the set of versions containing the \textit{vic} is [\texttt{binutils-2\_35}, \texttt{binutils-2\_42}], while the set of patched versions containing the \textit{vfc} is [\texttt{binutils-2\_36}, \texttt{binutils-2\_42}], with \texttt{binutils-2\_42} being the latest version of the Binutils project.
Therefore, we consider the version tags [\texttt{binutils-2\_35}, \texttt{binutils-2\_36}) as vulnerable. \hl{This notation signifies a range that includes \texttt{binutils-2\_35} but excludes \texttt{binutils-2\_36}}.

\section{Experimental Setup}

\subsection{Dataset}
With confirmation from the author of V-SZZ, it is established that V-SZZ assumes the deleted lines in the \textit{vfc} as vulnerable codes and manually labels the first introduction of deleted lines in a \textit{vfc} as the \textit{vic} to construct the dataset. However, as mentioned in Section~\ref{sec:motivating}, deleted lines cannot fully capture the vulnerability logic. Additionally, it cannot be guaranteed that the identified software version will necessarily exhibit the vulnerability.
Meanwhile, V-SZZ excludes the \textit{vfc}s that only contain added lines of code in their vulnerability selection, as the SZZ algorithm relies on tracking deleted lines of code to locate \textit{vic} and thus cannot handle such cases.
Therefore, we need to construct a ground-truth dataset with labeled vulnerable versions or non-vulnerable versions of OSS, which are verified by a public PoC.

\vspace{4px}
\noindent\textbf{Vulnerabilities.} 
To build a reliable evaluation dataset, we formulate four vulnerability selection criteria: 
1) The vulnerability must be a CVE reported in an OSS project;
2) The vulnerability must have publicly available PoCs and patches;
3) The PoC must be reproducible across different versions of the OSS and
4) The OSS must have a certain number of versions for comprehensive analysis.
Based on the selection criteria, we established a dataset comprising \hl{122} publicly disclosed CVEs associated with public PoC and patches, spanning 12 prevalent OSS, as detailed in Table~\ref{74CVEs}. 
The published date of these CVEs from 2016 to \hl{2025}.
%The selected OSS is widely used and has a certain number of versions. 
Meanwhile, the collected vulnerabilities cover 13 common CWE types. %as indicated in Figure~\ref{fig:cwe}. 
Some vulnerabilities are classified into multiple CWEs. For example, CVE-2023-1579 belongs to both CWE-119 (Improper Restriction of Operations within the Bounds of a Memory Buffer) and CWE-787 (Out-of-bounds Write). Among the dataset, vulnerabilities belonging to CWE-787 (Out-of-bounds Write) are the most numerous, accounting for 20.27\%.

\begin{table}[tp]
	\caption{Ground-truth Dataset Overview.}
	\begin{center}
		\begin{threeparttable}
			\begin{tabular}{l l l l l}
				\toprule
				\textbf{IDX} & \textbf{Name} & \textbf{\#Version} &\textbf{\#CVE} & \textbf{Domain} \\
				\midrule
				1 & Binutils & 157 & 18 & Programming tools\\
                2 & cJSON & 48 & 3  & JSON parser\\
                3 & FFmpeg & 405 & 31 & Multimedia processing \\
                4 & Jasper & 99 & 12 & Image coding toolkit \\
                5 & Libarchive & 51 & 10 & Streaming processing\\
                6 & Libcaca & 7 & 3 & Graphics library\\
				7 & Liblouis & 69 & 6 & Braille translator\\
				8 & Libming & 14 & 9 & Flash library\\
				9 & Libtiff & 78 & 7 & Image tools\\
                10 & Libxml2 & 233 & 12 & XML toolkit\\
				11 & OpenJPEG & 25 & 8 & Image codec\\
				12 & Pcre2 & 25 & 3 & Regular expression engine\\
				\midrule
				\textbf{Total} & - & \textbf{1,211} & \textbf{122} & - \\
				\bottomrule
			\end{tabular}
		\end{threeparttable}
	\end{center}
	\label{74CVEs}
\end{table}

% \iffalse
% \begin{figure}
%     \centering
%     \includegraphics[width=1.05\linewidth]{figs/cve_cutoff_analysis.png}
%     \caption{Temporal Distribution of CVEs with Knowledge Cutoff Boundary}
%     \label{fig:cutoff}
% \end{figure}
% \fi

%\noindent\hl{\textbf{Knowledge Cutoff Analysis.} The three LLMs in our evaluation have different knowledge cutoff dates: GPT-4's knowledge cutoff is October 2023, CodeLlama's training data extends to early 2023, and DeepSeek-V3's knowledge cutoff is July 2024. To ensure comprehensive evaluation across all models, we use DeepSeek-V3's cutoff date (July 2024) as the boundary since it represents the latest knowledge cutoff among the three models. As illustrated in Figure}~\ref{fig:cutoff}, \hl{our dataset includes 74 CVEs published before the knowledge cutoff and 48 CVEs published after the cutoff. This distribution enables us to evaluate \textsc{Vercation}'s performance on both vulnerabilities that may be known to the LLM and those that are completely unknown to all models.}

\vspace{4px}
\noindent\textbf{Patches.} We acquired security patches by crawling \textit{vfc} from the OSS Git repositories. In total, the \textit{vfc}s cover \hl{2,287} modification lines containing \hl{1,532} insertion lines and \hl{755} deletion lines. Among these, \hl{289} lines are unrelated to vulnerability (e.g., modifications in the ChangeLog file). 
%The distribution of the number of modification lines per patch is shown in \todo{Figure~\ref{fig:lines}}. 
On average, each \textit{vfc} has 21.8 modification lines. %As show in Figure~\ref{fig:mod_lines}, 
The number of added lines significantly exceeds that of deleted lines, underscoring the critical importance of incorporating additional code into the analysis process. Among these, 10.3\% are large \textit{vfc}s (the number of modification lines exceeding 50). Notably, 79 \textit{vfc}s (64.75\%) contain both code insertion and deletion, 29 \textit{vfc}s exclusively insertions (23.77\%), and 15 \textit{vfc}s solely involve deletions (12.30\%). 

% \iffalse
% \begin{figure}
%     \centering
%     \includegraphics[width=0.9\linewidth]{figs/lines.png}
%     \caption{The average distribution of modification lines in fixing commits for each OSS.}
%     \label{fig:mod_lines}
% \end{figure}

% \begin{figure}
%     \centering
%     \includegraphics[width=0.9\linewidth]{figs/CWE.png}
%     \caption{The CWE Type Distribution.}
%     \label{fig:cwe}
% \end{figure}
% \fi

\vspace{4px}
\noindent\textbf{Verifying \& Labeling.}
We apply the following process to systematically label the ground-truth dataset:

\begin{enumerate}[label=(\roman*),leftmargin=*]
\item The fixed version can be determined through the official release of \textit{vfc}s. We confirm whether versions following the fixed version are vulnerable by checking if they contain the patch code.

\item For the remaining version, we obtained the public PoC of each vulnerability from the Git issues or by referring to OSS sites. 

\item To validate the PoC on each corresponding OSS version, we establish the dependent environment for each OSS version and compile the libraries.

\item %During the PoC testing, we observe whether vulnerabilities are triggered on $S_v$. If they can be triggered, we mark these statements as $S_trigger$ and label the version as the vulnerable version. For versions that cannot be triggered by the PoC (some PoCs only apply to specific versions), we check if $S_trigger$  is included in the version to label the version accordingly. %在poc测试过程中，我们观察是否在crucial vulnerable statements上触发漏洞。如果可以触发，则将这些statements标记为Sp并将version label为vulnerable version. 对于Poc无法触发漏洞的版本（因为一些poc只适用于部分特定版本），我们检查version中是否包含Sp来为version打上标签。
During the PoC input testing process for each vulnerability, we analyze the triggering conditions and dangerous behaviors of the vulnerability and label the statements that trigger the vulnerability as \var{S_{trigger}}. For versions where the PoC cannot trigger the vulnerability (as some PoCs are only applicable to certain specific versions), we check whether the version contains \var{S_{trigger}} to label the version.
\end{enumerate}
In our experimental setup, we took 10 days to build the software and version pool. Then we took 30 days to build the ground-truth dataset, including the time taken to establish the dependent environment for each software version, as well as the time to perform vulnerability verification and labeling.

\subsection{Baseline and Metrics}
\noindent\textbf{Baseline.} For the vulnerable version identification task, we selected the NVD~\cite{CPE}, along with SOTAs: SZZ-based methods (AG-SZZ~\cite{szz_kim2006automatic}, B-SZZ~\cite{B-SZZ}, V-SZZ~\cite{bao2022v}) and clone-based methods (V0Finder~\cite{woo2021v0finder}, V1SCAN~\cite{clone_woo2023v1scan},  VERJava~\cite{sun2022verjava} and Vision~\cite{wu2024vision}) for accuracy comparison. We gathered the vulnerable version range of each CVE from CPE to evaluate the NVD's accuracy. 

The basic idea of SZZ algorithms is to backtrack the commit history to locate the earliest commit that introduces the deletion lines removed by security patches. We use different SZZ algorithms to identify the \textit{vic}, with the versions between the \textit{vic} and the \textit{vfc} considered vulnerable. Since these algorithms trace deleted lines individually, instances may arise where multiple \textit{vic} are identified, and in such cases, we select the earliest one as the detection outcome.

%V0Finder is a vulnerability introduction detection tool that uses a clone-based technique. We first use V0Finder to generate vulnerable function fingerprints for our ground-truth dataset, and then generate fingerprints for all functions of all OSS versions. Subsequently, we use V0Finder's clone-based approach to identify vulnerable versions.}

%\sethlcolor{light-gray}\hl{V1SCAN is an approach that combines version-based and code-based approaches for discovering 1-day vulnerabilities in reused C/C++ OSS components. It first classifies reused code into exactly reused, changed and unused categories. Then it uses both version identification and code clone detection to identify vulnerabilities. We use V1SCAN's technique to detect the code clone for each version and identify the affected versions of the vulnerabilities in the ground-truth dataset.}

%\sethlcolor{light-gray}\hl{VERJava is specifically designed for Java OSS and employs a two-stage analysis approach. In the first stage, it performs function-level vulnerability existence analysis. The second stage analyzes the existence of the vulnerability in OSS versions at the entire patch level by considering the proportion of patched functions.} %While effective for Java projects, its analysis granularity at function level may miss some vulnerability patterns that exist at statement level."

V0Finder, V1SCAN, VERJava, and Vision are clone-based methods. We use their techniques to detect code clone for each version and identify the affected versions of the vulnerabilities in the ground-truth dataset. Because VERJava and Vision are applied in Java, we modified the code to make them compatible with C/C\texttt{++}.

For \textsc{Vercation}, \hl{we} selected thresholds $\vartheta_1=0.9$, $\vartheta_2=0.8$ and $\vartheta_3=0.7$ (related experiments are introduced in Section~\ref{sec:eval_rq3_code_clone}).
In terms of the performance of vulnerability code extraction, we compared the differences between combining various LLMs (GPT-4, CodeLlama-13B, DeepSeek-V3) and using the static analysis tool Joern parser~\mbox{\cite{Joern}} alone.
%We used SZZ algorithms and V0Finder with its default options by referring to the paper~\cite{szz_kim2006automatic, B-SZZ, bao2022v, woo2021v0finder}. 

\vspace{4px}
\noindent\textbf{Metrics.} To evaluate the accuracy, we employ the following three metrics, i.e., true positives (\var{TP}), false positives (\var{FP}), false negatives (\var{FN}), precision ($\frac{\#TP}{\#TP+\#FP}$), recall ($\frac{\#TP}{\#TP+\#FN}$) and F1-score ($\frac{2*precision*recall}{precision+recall}$), to measure the accuracy of the above methodologies. These metrics are also used in previous studies to evaluate the technique's performance~\cite{bao2022v,rosa2021evaluating}. 
Because of significant differences in the version count for distinct OSS, we compute the precision and recall for each OSS and report the average of precisions and recalls finally.

\subsection{Implementation Details}
We utilized Joern~\cite{Joern} for program slicing and developed Python scripts to extract control and data dependencies, outputting the sliced statements as dangerous flows.
We explored three LLM models, GPT-4~\cite{gpt4}, CodeLlama-13B~\cite{roziere2023codellama}, and DeepSeek-V3~\cite{deepseekai2024deepseekv3technicalreport} for vulnerable code extraction. 
We use the public API developed by OpenAI to perform the experiment in GPT-4. The API version is \texttt{GPT-4-0613} published in 2023. 
CodeLlama was created through further fine-tuning of Llama 2 on specific code datasets. 
We selected the 13B parameter model (\texttt{codellama/CodeLlama-13b-hf}) from the CodeLlama model series. The API of DeepSeek-V3 was published in July 2024. \hl{We set the temperature parameter to 1.0. To account for the stochastic nature of LLM outputs, we generate 10 independent responses for each CVE}~\cite{sallou2024breaking}. \hl{We employ a majority voting strategy where statements appearing in $\geq$ 6 out of 10 runs are included in the final vulnerable statement set for evaluation.} The maximum size of tokens is 1,024. %In all our experiments, we use the top-1 prediction from the model.
We deployed the Codellama model on our server with four NVIDIA RTX A5000 GPUs. Similar to GPT-4, the temperature parameter is set to 1.0, and the maximum size of tokens is set to 1,024.
Subsequently, we apply weight allocation strategies to extract vulnerable statements using Python scripts. \hl{For threshold selection, we used $\theta_1$ = 0.9, $\theta_2$ = 0.8 and $\theta_3$ = 0.7 based on 
grid search optimization described in our sensitivity analysis.}
In the code change detection, we utilize the Clang 10.0.0 tool for function expansion (inline optimization) and AST generation of C/C\texttt{++} code. Specifically, we use the \textit{-fsyntax-only} option to perform syntax checking without actual compilation. We then develop Python scripts to achieve AST normalization and similarity comparison. Overall, we constructed our system with 6K LoC in Python. 
\section{Evaluation}
In this section, we evaluate the performance of the proposed \textsc{Vercation} by answering the following research questions (RQs).

\begin{itemize}[leftmargin=*]
\item\textbf{RQ1.} \textit{Overall Effectiveness:} How does the overall performance of \textsc{Vercation} compare against state-of-the-art vulnerable version identification methods?
\item\textbf{RQ2.} \textit{Architectural Contribution:} How do the key architectural choices in \textsc{Vercation}—the use of an LLM, static analysis, and statement weighting—contribute to its performance?
\item\textbf{RQ3.} \textit{Component Robustness:} How accurate and robust is \textsc{Vercation}'s statement-level code clone detection?
\item\textbf{RQ4.} \textit{Real-World Application:} What are the effects of applying \textsc{Vercation} in the real world scenarios?
\end{itemize}

\subsection{RQ1: Overall Effectiveness}
We first compare \textsc{Vercation} against the NVD and SOTA methods.
Table~\ref{tab:comparison_sota} shows the vulnerable version identification results of NVD, SZZ algorithms, V0Finder, V1SCAN, VERJava, Vision, and \textsc{Vercation}. 
Note that some CVEs do not provide vulnerable versions in NVD, such as CVE-2021-30499~\cite{CVE-2021-30499}.
SZZ algorithms cannot work on the \textit{vfc}s containing solely added lines (i.e., 11 cases), so we exclude these cases in approaches using the SZZ algorithm.
% \zt{Does this mean only SZZ algorithms were evaluated on test set - 11 cases; all the rest methods were evaluated on the whole test set?}
% Meanwhile, we evaluated different LLM models and prompt strategies in \textsc{Vercation}'s phase 1 to select the best solution.

\subsubsection{Comparison with NVD}
The NVD achieves a precision of 66.8\% and a recall of 41.8\%.
False positives often occur because NVD may broadly flag all versions preceding a patch as vulnerable without specific analysis. 
This situation exists in 18 CVEs (15\%) of the ground truth.
For instance, security experts found a buffer overflow risk exists in \texttt{Libtiff 4.4.0}, but NVD simply reported the versions of \texttt{Libtiff$<$4.4.0} were exposed in vulnerability. In reality, the vulnerable version range is \texttt{Libtiff}$<$4.4.0 and $\ge$4.0.0. 

The reason for low recall in NVD can be summarized into two types: 
(i) NVD only reported the version in which vulnerabilities were found. For example, CVE-2017-14152~\cite{CVE-2017-14152} was found in \texttt{OpenJPEG} 2.2.0, then NVD reported the affected version was only 2.2.0, whereas the real vulnerable version range was 2.2.0-2.2.1. The problem exists in 57 CVEs (47\%) of the ground truth. 
(ii) Some vulnerabilities are not fixed promptly after disclosure but are addressed after several versions have been released. These vulnerable versions can be overlooked if NVD does not update the version information. 
For example, CVE-2021-33815~\cite{CVE-2021-33815} was discovered in \texttt{FFmpeg} 4.4 but was fixed in version 5.0. Due to the lack of timely tracking of the fix information, NVD overlooked the vulnerable versions 4.4.1-4.4.4.

\begin{table}[tp]
    \centering
    \caption{Comparison with NVD Database and SOTA works.}
    \begin{threeparttable}
    \begin{tabular}{l  l c c c}
    \toprule
     \textbf{Type} & \textbf{Methods}  & \textbf{Precision} & \textbf{Recall} & \textbf{F1 score}  \\
     \midrule
      - & NVD Database & 0.668 & 0.418 & 0.514 \\
     \midrule
     \multirow{3}{*}{\makecell{SZZ\\Algorithm}}
     ~ & AG-SZZ~\cite{ag-szz} & 0.372 & 0.625 & 0.466\\
     ~ & B-SZZ~\cite{B-SZZ} & 0.378 & 0.543 & 0.446\\
     ~ & V-SZZ~\cite{bao2022v} & 0.756 & 0.851 &0.801\\ 
     \midrule
     \multirow{4}{*}{\makecell{Clone-based\\Approach}}
     ~ & V0Finder~\cite{woo2021v0finder} & 0.829 & 0.745 & 0.785\\
     ~ & V1SCAN~\cite{clone_woo2023v1scan} & 0.863 & 0.724 & 0.788\\
     ~ & VERJava~\cite{sun2022verjava} & 0.891 & 0.247 & 0.386\\
     ~ & Vision~\cite{wu2024vision} & 0.842 & 0.881 & 0.861\\
     \midrule
     - & \textsc{Vercation} & \textbf{0.918} & \textbf{0.945} & \textbf{0.931}\\
    \bottomrule
    \end{tabular}
    \end{threeparttable}
    \label{tab:comparison_sota}
\end{table}

\subsubsection{Comparison with SZZ-based Methods}
The basic idea of SZZ algorithms is to backtrack the commit history to locate the earliest commit that introduces the deletion lines removed by patches. 
Regarding F1-score, \textsc{Vercation} improves the best-performing baseline V-SZZ by 16.2\%. And the difference between the F1 of B-SZZ and AG-SZZ is small (0.446 vs. 0.466).
We affirmed that there are two main causes of false alarms in these approaches: 

\begin{enumerate}[label=(\roman*),leftmargin=*]
\item The line mapping algorithm based on edit distance does not consider the behavior of the source code. As shown in Listing~\ref{semantic}, 
The deleted lines were encapsulated within the added functions, and in fact, they have the same behavior. Line mapping algorithms failed in these cases, unable to identify the true \textit{vic}. %example OpenSSL 747e16398d7
\item The SZZ algorithm treats each deleted line as an independent origin for backtracking, resulting in multiple \textit{vic} being identified. For example, the security patch of CVE-2020-35965~\cite{cve-2020-35965} from \texttt{FFmpeg} has 5 deleted lines. 
The true vulnerability logic lies in the lack of checking the size of \texttt{ymax} before executing the \texttt{memset} zero operation, potentially resulting in out-of-bounds writes to memory. Our method utilizes a weighted allocation approach to increase the weight of \texttt{memset} operation, thus backtracing to the correct \textit{vic}. However, SZZ algorithms separately backtrack different deleted lines, resulting in the identification of 5 different \textit{vic}, significantly reducing the precision. 
\end{enumerate}

\subsubsection{Comparison with Clone-based Methods}
In the same dataset, V0Finder achieved a precision of 82.9\% and a recall of 74.5\%. There are two reasons for the inaccuracies of V0Finder: 1) V0Finder generates a hash value for the entire vulnerable function as a vulnerability fingerprint. This fingerprint is coarse because the vulnerable function contains a lot of vulnerability-irrelevant code, introducing a significant amount of noise. 2) V0Finder considers the function containing all deleted patch code as a vulnerable function clone. This approach also overlooked the inserted code of the security patch.

V1SCAN achieves an F1-score of 0.788, combining version-based detection with code-based detection methods.
For version-based detection, V1SCAN obtains CVE-affected OSS versions from NVD's CPE database and filters out functions belonging to versions outside the CVE's affected range. As we mentioned earlier, there are many errors in NVD's affected version information, which consequently affects V1SCAN's detection results.
For code-based detection, V1SCAN primarily relies on text similarity comparisons, overlooking deeper relationships between code snippets. When functions are refactored or outlined into new functions, this approach fails to identify vulnerability-related code, resulting in false negatives.

Table~\ref{tab:comparison_sota} reports a low recall of 0.247 in VERJava. VERJava has a fundamental issue in determining vulnerable versions: it requires the target version's code to simultaneously meet two strict conditions - perfect matching of deleted lines in the \textit{vfc} (delSim $\ge$ 1.0) and the absence of most added patch code (addSim $\le$ 0.9). In project development, code frequently undergoes changes, including variable name modifications and function refactoring. This means that even versions that contain vulnerabilities are unlikely to meet such strict matching requirements due to code evolution. As a result, many vulnerable versions are incorrectly identified as safe versions, leading to serious false negatives.
Another problem with VERJava is similar to V0Finder, it incorrectly assumes that deleted lines in the patch represent vulnerable features.

Vision achieves an F1 of 0.861, combining critical statement selection with weighted dependency graphs. \hl{Vision relies on taint analysis as a first step for critical statement selection. However, this approach often includes many statements that are control or data-dependent but not actually relevant to the vulnerability, which consequently affects Vision's precision in extracting vulnerable code.} Moreover, Vision's analysis is limited to Java packages from Maven repositories, which may not fully capture the vulnerability patterns in other languages.

\vspace{4px}
\noindent\textbf{Accuracy of \textsc{Vercation}.}
\textsc{Vercation} demonstrates consistent performance across most projects, achieving high precision ($>$ 0.90) on 9 out of 12 projects and perfect recall (1.00) on several projects, including cJSON and Libcaca. Notably, while baseline methods like V-SZZ and V0Finder show significant performance fluctuations across different projects, \textsc{Vercation} maintains more stable performance. For instance, on the Binutils project, which has complex version management and frequent refactoring, Vercation achieves a 0.95 F1 score, outperforming V-SZZ (0.47 F1 score) and V0Finder (0.88 F1 score).

Though \textsc{Vercation} performed well in the version identification, it still failed to identify inducing commits for a small number of vulnerabilities. 
One reason is that in some cases, slight modifications in the AST have little impact on the similarity between the AST before and after the modification, but are crucial in introducing vulnerabilities.
Another reason is that despite using the weight allocation strategies, there are still some statements related to patch variables that are not relevant to the vulnerability but have been assigned high weights.
The above two reasons may incorrectly identify \textit{vic}, leading to a misidentification of the vulnerable version range.

\hl{In our evaluation, 5.7\% of traced commit histories involved merge commits with multiple parents. For example, in CVE-2022-1355}~\cite{cve-2022-1355} \hl{from Libtiff, the vulnerability was introduced in a feature branch that was later merged into the main development line, creating a non-linear history with two parent commits at the merge point. For these cases involving non-linear histories, \mbox{\textsc{Vercation}} achieved a precision of 91.2\% and recall of 92.5\%, demonstrating that our multi-path traversal strategy effectively handles branching scenarios without significant performance degradation.}

\begin{center}
\fcolorbox{black}{gray!10}{\parbox{0.96\linewidth}{\textbf{RQ1}: \textsc{Vercation} demonstrates superior performance in vulnerable version identification on our ground-truth dataset, achieving an F1 score of 93.1\%. This significantly outperforms SOTA works. The prompt strategy combining Few-shot and CoT techniques performs the best.}}
\end{center}

\subsection{RQ2: Architectural Contribution}
\label{sec:evaluation_llm}

\subsubsection{Program Slicing \& LLM Ablation Study}
We conducted ablation experiments by separately removing the program slicing component (implemented by Joern) and the LLM component (DeepSeek) from phase 1.

\begin{figure*}[htbp] % Add float placement specifiers
    \centering
    
    \begin{subfigure}[t]{0.32\textwidth}
        \centering
        \includegraphics[width=\linewidth]{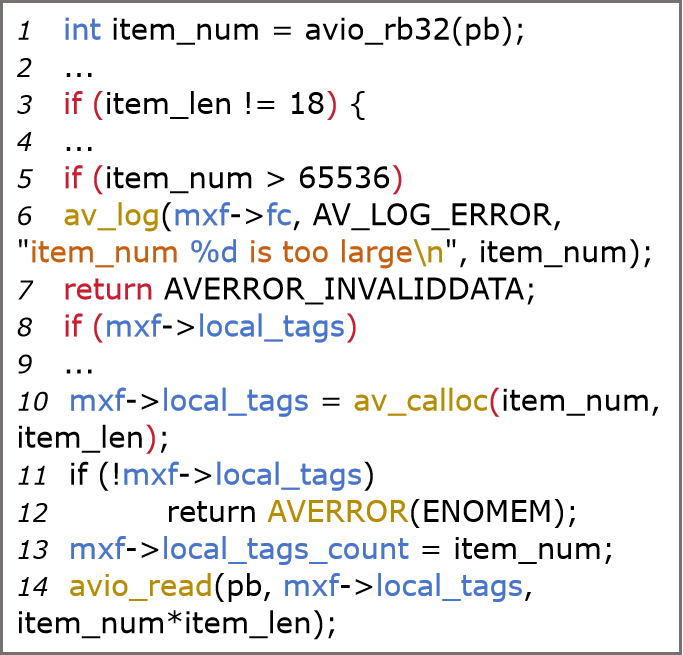}
        \caption{Extracted by Using Only Joern Parser.}
        \label{fig:use_joern}
    \end{subfigure}%
    \hfill 
    \begin{subfigure}[t]{0.32\textwidth}
        \centering
        \includegraphics[width=\linewidth]{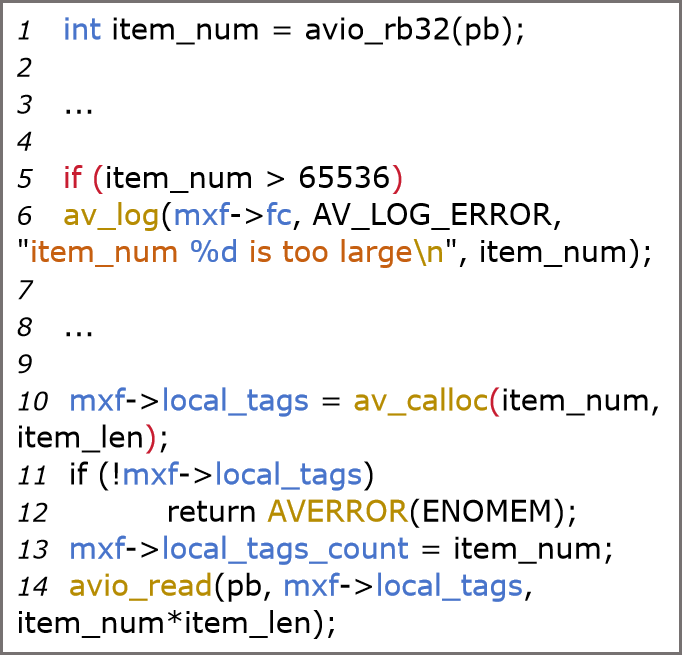}
        \caption{Extracted by Using Only LLM.}
        \label{fig:use_llm}
    \end{subfigure}%
    \hfill
    \begin{subfigure}[t]{0.32\textwidth}
        \centering
        \includegraphics[width=\linewidth]{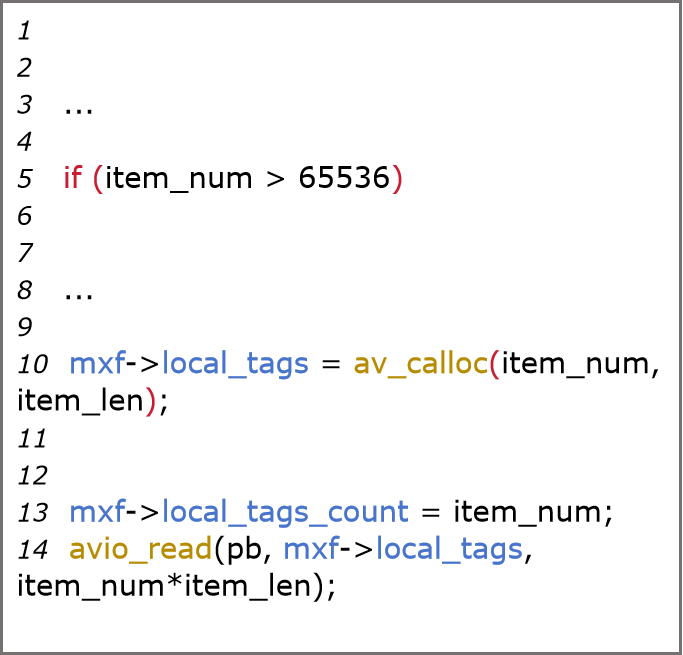}
        \caption{Extracted by Combining Joern Parser and LLM.}
        \label{fig:use_joern_llm}
    \end{subfigure}
    
    \caption{The Vulnerable Statements Extracted by Different Methods.}
    \label{fig:sample_subfigures}
\end{figure*}

\begin{table}[tp]
    \centering
    \caption{Performance of LLMs in Vulnerability Logic Comprehension.}
    \label{tab:llm_result}
    
    \begin{threeparttable}
        \begin{tabular}{@{} l rrrr @{}} % Use booktabs-friendly specifiers
            \toprule
            \textbf{Method (Tool)} & \textbf{Avg \#Vuln Stmts}\tnote{1} & \textbf{Precision} & \textbf{Recall} & \textbf{F1} \\
            \midrule
            \makecell[l]{Static Analysis (Joern)} & 22.64 & 0.435 & 0.544 & 0.483 \\
            \midrule
            \addlinespace % Adds a little extra space between rows
            \makecell[l]{LLM  (DeepSeek)} & 11.33 & 0.742 & 0.796 & 0.768 \\
            \midrule
            \makecell[l]{Static Analysis + LLM \\ (Joern + DeepSeek)} & 5.66 & \textbf{0.918} & \textbf{0.945} & \textbf{0.931} \\
            \bottomrule
        \end{tabular}
        
        \begin{tablenotes}
            \footnotesize
            \item[1] The average count of extracted vulnerable statements for each CVE.
        \end{tablenotes}
    \end{threeparttable}
    
\end{table}

\vspace{4px}
\noindent\textbf{Static Analysis Only.}
As depicted in Table~\ref{tab:llm_result}, the effectiveness of Joern in extracting vulnerability logic is significantly inferior to that harnessed by the power of LLM models, with an F1 score 46.9\% less than Joern + DeepSeek (Few-shot + CoT). This discrepancy primarily stems from Joern's reliance on fixed dependency extraction patterns, leading to the retrieval of numerous statements irrelevant to vulnerabilities. 
As shown in Figure~\ref{fig:use_joern}, using only the static analysis tool Joern parser for program slicing extracts statements on Lines 3, 7-8, 11-12. While these statements are related to control flow, they are not directly relevant to the vulnerability. This illustrates how static analysis alone can include extraneous information that is not crucial to understanding the specific vulnerability logic.
Meanwhile, we found that Joern cannot accurately address pointer structure dataflows, thus missing some statements that have data or control dependencies. To elaborate, on average, Joern extracts 3.64 times more vulnerable statements per CVE compared to combined DeepSeek. 

\vspace{4px}
\noindent\textbf{LLM Only.}
The approach of LLM Only (DeepSeek) involves inputting the entire function into the model for analysis. As illustrated in Table~\ref{fig:use_llm}, DeepSeek alone (with Few-shot + CoT strategies) achieves an F1 score of 0.739, which is 18.7\% lower than the Joern + DeepSeek combination (0.926). This performance gap primarily stems from LLM's tendency to include a broader range of contextual information when given an entire function as input, this over-inclusion of context leads to lower precision because some of which are tangential to the core vulnerability logic. 
As illustrated in Figure~\ref{fig:use_llm}, using only LLM improves upon static analysis by reducing the extraction of irrelevant control flow statements. However, when given the entire function as input, the LLM struggles to precisely differentiate between core vulnerability-related code and contextual information. This results in the inclusion of some noise, such as variable definitions (Line 1) and unrelated control flow statements (Lines 4, 7-8). %The LLM's broad contextual understanding, while powerful, can sometimes lead it to overinclude information it deems potentially relevant, even if not directly tied to the vulnerability.

As shown in Figure~\ref{fig:use_joern_llm}, the combination of Joern's structured analysis with DeepSeek's code understanding allows for a more focused and accurate extraction of vulnerability-related statements. 
Joern provides an initial set of potentially relevant statements based on program structure, which DeepSeek then refines using its contextual understanding, resulting in a more balanced and effective approach to vulnerability logic comprehension.

\subsubsection{LLM Performance, Consistency, and Generalization} We further evaluated the LLM component by comparing different models and analyzing their stability and ability to generalize.

\begin{table}[tp]
    \centering
    \caption{\textsc{Vercation} with Different Strategies.}
    \begin{threeparttable}
    \begin{tabular}{l l rrr}
    \toprule
    \textbf{Model} & \textbf{Strategy} & \textbf{Precision} & \textbf{Recall} & \textbf{F1 score}\\
    \midrule
    \multirow{3}{*}{GPT-4} & Zero-shot  & 0.708 & 0.823 & 0.761\\
     & Few-shot  & 0.892 & 0.907 & 0.899\\
     & Few-shot + CoT   & 0.893 & 0.946 & 0.925\\
     \midrule
              \multirow{3}{*}{CodeLlama} & Zero-shot   & 0.671 & 0.789 & 0.725\\
              & Few-shot   & 0.851 & 0.895 & 0.873\\
              & Few-shot + CoT   & 0.842 & 0.912 & 0.876\\
              \midrule
              \multirow{3}{*}{DeepSeek-V3} & Zero-shot   & 0.769 & 0.827 & 0.797\\
               & Few-shot   & 0.882 & 0.921 & 0.901\\
               & Few-shot + CoT   & \textbf{0.918} & \textbf{0.945} & \textbf{0.931}\\
   \bottomrule
    \end{tabular}
    \end{threeparttable}
    \label{tab:comparison_strategy}
\end{table}

\vspace{4px}
\noindent{\textbf{Comparative Performance.}} Table~\ref{tab:comparison_strategy} illustrates the performance of combining the dangerous flows extracted by the Joern parser with LLMs on our dataset, with the ``Strategy'' column delineating different prompting strategies.
Our observation reveals that DeepSeek-V3 prompted with Few-shot and CoT surpasses other models. 
Specifically, DeepSeek-V3 (Few-shot + CoT) exhibits a 0.64\% and 6.28\% improvement in F1-score compared to the GPT-4 and CodeLlama, respectively. Moreover, the performance trends with DeepSeek remain consistently upward across different prompting strategies. When prompted with the Few-shot prompt, DeepSeek achieves a 13.15\% improvement in F1-score over the Zero-shot prompt. This is further improved by 3.33\% F1 with the addition of CoT prompting. 
We observed that the utilization of the Zero-shot prompt often results in an increased generation of false positives (vulnerable statements), including the statements predefining patch variables (e.g., Line 6 in Listing~\ref{list:CVE-2017-14169}). Conversely, in examples prompted by Few-shot, we prioritize understanding the triggering mechanisms of vulnerabilities, thereby facilitating LLM in learning a more accurate method of generating vulnerability logic. 

\begin{table}[tp]
    \centering
    \caption{\protect\hl{LLM Consistency for Vulnerable Statement Extraction.}}
    \label{tab:llm_consistency}
    \begin{tabular}{lcc} 
        \toprule % Top line
        \textbf{Model} & \textbf{Agreement Rate} & \textbf{Jaccard Similarity} \\
        \midrule % Middle line
        GPT-4          & 0.454                   & 0.439                   \\
        \midrule
        CodeLlama      & 0.517                   & 0.566                   \\
        \midrule
        DeepSeek-V3    & \textbf{0.586}          & \textbf{0.629}          \\
        \bottomrule % Bottom line
    \end{tabular}
    
\end{table}

\vspace{4px}
\noindent\hl{\textbf{LLM Consistency Analysis.} To evaluate the stability of our LLM-based vulnerable statement extraction, we measured two consistency metrics across 10 independent runs with temperature = 1.0. 
The agreement rate represents the percentage of statements that appeared in all 10 runs relative to the total unique statements extracted across all runs. As shown in Table}~\ref{tab:llm_consistency} \hl{DeepSeek-V3 achieved the highest agreement rate 58.6\%, followed by CodeLlama (51.7\%). 
For cases with variation, we calculated average Jaccard similarity by computing pairwise similarities between all run combinations and taking the mean. DeepSeek-V3 achieved the highest Jaccard similarity of 0.629 shown in Table}~\ref{tab:llm_consistency}, \hl{indicating that DeepSeek-V3 maintains the most consistent performance even when exact matches were not achieved. 
While the consistency levels are moderate across all models, this reflects the inherent complexity of vulnerability logic analysis where slight variations in reasoning can lead to different but potentially valid interpretations. 
The superior consistency of DeepSeek-V3, combined with its highest F1-score performance, makes it the most reliable choice for practical deployment.}

\begin{figure*}[tp]
    \centering
    \includegraphics[width=0.75\linewidth]{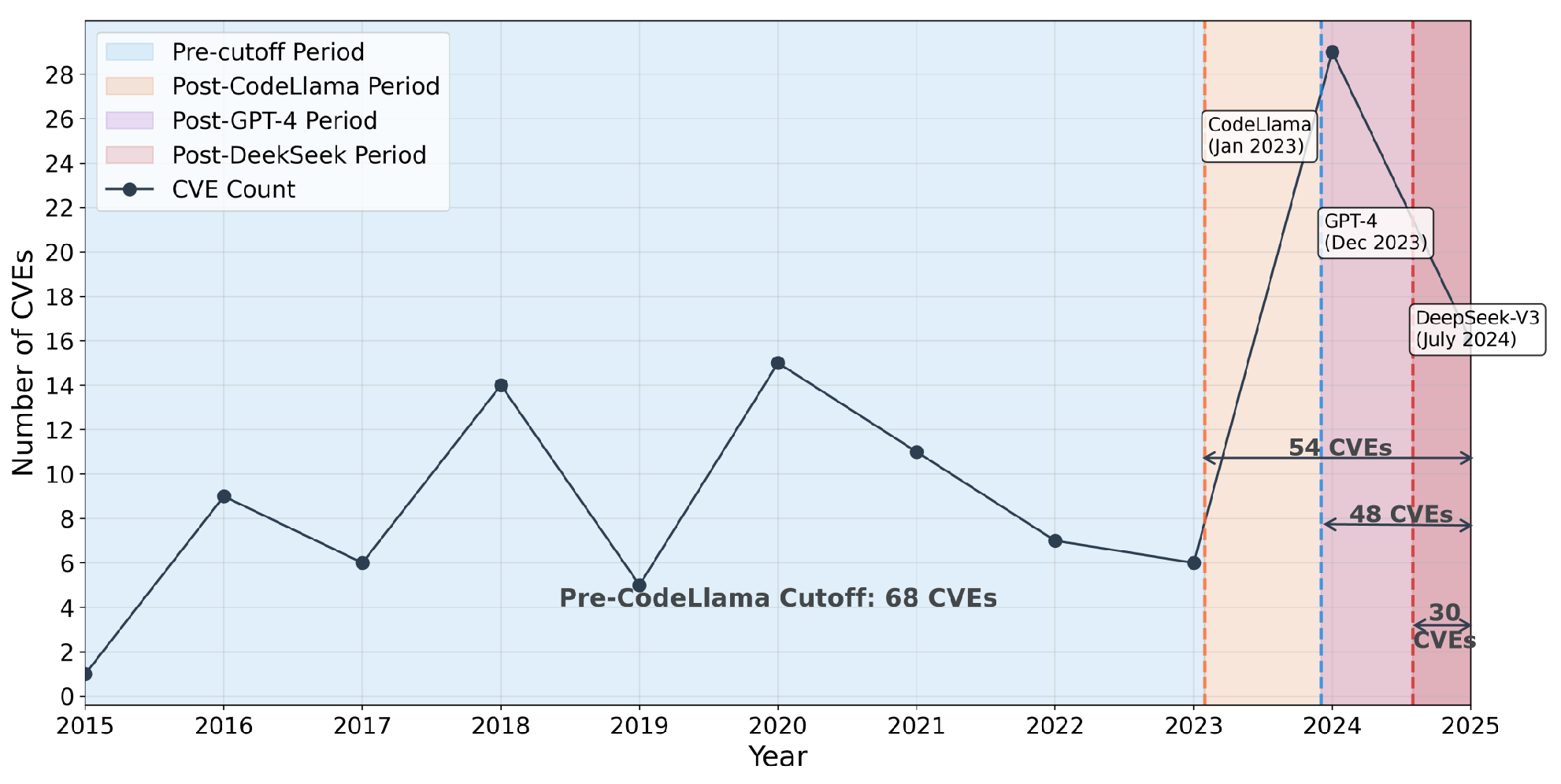}
    \caption{Temporal Distribution of CVEs with Knowledge Cutoff Boundary.}
    \label{fig:cutoff_all}
\end{figure*}

\vspace{4px}
\noindent{\textbf{Generalization Capability.}}
\hl{
A critical concern regarding the LLM-based approach is whether the models' knowledge cutoff affects performance when analyzing vulnerabilities published after the training data cutoff. 
To address this concern and demonstrate that \textsc{Vercation} leverages LLMs' general code understanding capabilities rather than specific vulnerability knowledge, we conducted an ablation study comparing each model's performance on CVEs published before and after their respective knowledge cutoff dates.

The three LLMs in our evaluation have different knowledge cutoff dates: CodeLlama's training data extends to September 2022 (based on Llama 2's cutoff date}~\cite{touvron2023llama}), \hl{GPT-4's knowledge cutoff is December 2023}~\cite{gpt4cutoff}, \hl{and DeepSeek-V3's knowledge cutoff is July 2024 (confirmed through direct inquiry with DeepSeek). We partitioned our dataset according to each model's cutoff date and evaluated their performance on both pre-cutoff and post-cutoff CVE subsets as shown in Figure}~\ref{fig:cutoff_all}. 

\hl{As shown in Table}~\ref{tab:cutoff}, \hl{CodeLlama shows a slight performance decrease of 1.25\% (from 0.881 to 0.870) when analyzing post-cutoff CVEs. In contrast, both GPT-4 and DeepSeek-V3 demonstrate slight performance improvements on post-cutoff CVEs, with GPT-4 improving by 0.54\% and DeepSeek-V3 by 1.73\%. The performance variations across all models are minimal ($<$ 2\%), indicating remarkable consistency regardless of knowledge cutoff boundaries. These results validate our core hypothesis that \textsc{Vercation}'s effectiveness stems from leveraging LLMs' fundamental code understanding and reasoning capabilities rather than memorized vulnerability-specific knowledge.}

\definecolor{mydarkred}{RGB}{139,0,0}       % 深红色 (DarkRed)
\definecolor{mydarkgreen}{RGB}{0,100,0}     % 深绿色 (DarkGreen)

\begin{table}[tp]
    \centering
    \caption{Performance Comparison Across Knowledge Cutoff Periods.}
    \begin{tabular}{l l  l r}
      \toprule
       \textbf{Model}  & \textbf{Cutoff Date} &  \textbf{Period} & \textbf{F1-score} \\
       \midrule
         \multirow{2}*{CodeLlama} & \multirow{2}*{2022.09} & Pre-cutoff &  0.881\\ 
         \cmidrule{3-4}
         ~ & ~ & Post-cutoff & 0.870 (\textcolor{mydarkred}{$\downarrow$} 1.25\%) \\ 
       \midrule
         \multirow{2}*{GPT-4} &  \multirow{2}*{2023.10} & Pre-cutoff  & 0.923\\ 
         \cmidrule{3-4}
         ~ & ~ & Post-cutoff &  0.928 (\textcolor{mydarkgreen}{$\uparrow$} 0.54\%)\\ 
       \midrule
         \multirow{2}*{DeepSeek-V3} & \multirow{2}*{2024.07} & Pre-cutoff & 0.927\\ 
         \cmidrule{3-4}
         ~ & ~ & Post-cutoff & 0.943 (\textcolor{mydarkgreen}{$\uparrow$} 1.73\%)\\ 
        \bottomrule
    \end{tabular}
    \label{tab:cutoff}
\end{table}

\begin{figure}
    \centering
    \includegraphics[width=0.98\linewidth]{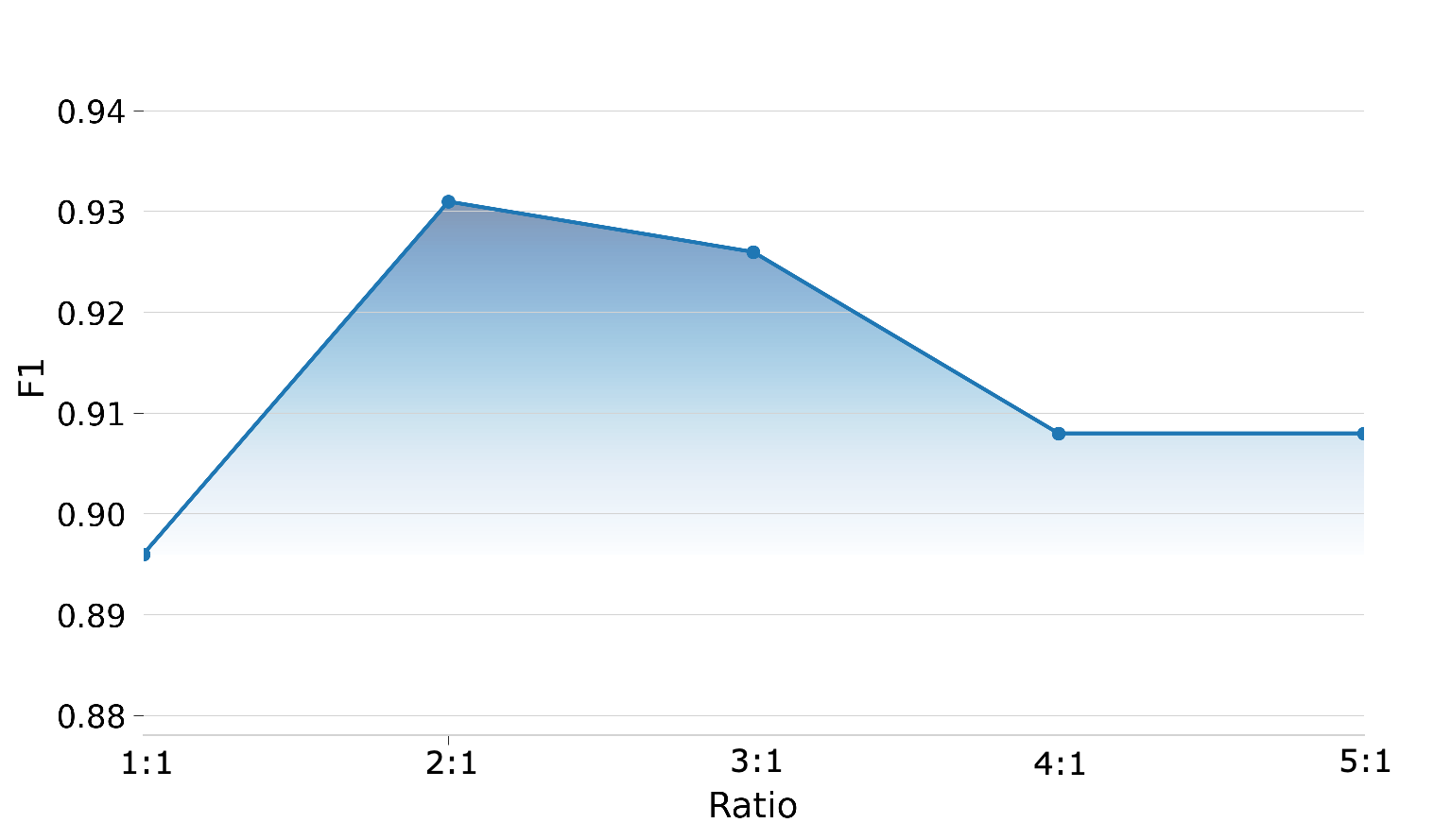}
    \caption{Impact of Statement Weight Ratios on F1-Score}
    \label{fig:ratio}
\end{figure}

\subsubsection{Effectiveness of the Statement Weighting Strategy}
\hl{To validate the effectiveness of our statement weighting strategy, we conducted an ablation study comparing our weighted approach with different weight ratios across our entire dataset.} 
\hl{We systematically evaluated weight ratios from 1:1 to 5:1, where the first number represents the weight for sensitive function statements ($W_{sensitive}$) and the second number represents the weight for regular statements ($W_{base}$). These ratios were tested using the DeepSeek-V3 model, with all other parameters held constant.}

\hl{Figure}~\ref{fig:ratio} \hl{illustrates the impact of different weight ratios on F1-score. The results demonstrate that equal weighting (1:1 ratio) achieves an F1-score of 0.896, while our proposed 2:1 ratio ($W_{sensitive}$ = 2.0, $W_{base}$ = 1.0) achieves the optimal performance of 0.931, representing a 3.91\% improvement. 
Performance gradually decreases as the weight ratio increases beyond 2:1. Higher ratios (3:1 and above) lead to over-emphasis on sensitive functions.
This causes the similarity score to be dominated by a small number of statements, reducing the method's ability to distinguish between true vulnerability patterns and coincidental sensitive function usage.}

\begin{center}
\fcolorbox{black}{gray!10}{\parbox{0.96\linewidth}{\textbf{RQ2}: 
\textsc{Vercation}'s effectiveness stems from the synergistic combination of static analysis and LLMs. DeepSeek-V3 with a Few-shot+CoT prompt proves to be the most accurate, consistent, and generalizable model. Furthermore, our 1:2 statement weighting strategy is shown to be optimal for balancing the influence of sensitive functions.
}}
\end{center}

\subsection{RQ3: Component Robustness}
\label{sec:eval_rq3_code_clone}
Here we evaluate the accuracy of the similarity comparison technique utilized in \textsc{Vercation}'s phase 2. 

\subsubsection{Accuracy of Code Clone Detection}
We compare the code clone detection technique in \mbox{\textsc{Vercation}}'s phase 2 with typical statement-level code similarity detection methods: Edit distance~\mbox{\cite{sheneamer2015distance}}, Hash value~\mbox{\cite{hummel2010hash}}, and CodeBERT embedding comparison~\mbox{\cite{sonnekalb2022codebert}}.
Current SOTA code clone detection methods primarily focus on function-level detection~\cite{fang2020functional}, whereas our method operates at the statement level. To perform a comparison, during the process of backtracing the previous commits of vulnerable statements, we apply code clone detection to the entire function before and after the modification. If the commit includes changes to other non-vulnerable statements within the function, we normalize these statements to be identical.
Table~\ref{tab:codeclone} summarizes the accuracy of different code clone detection methods. 

Edit distance~\cite{sheneamer2015distance} calculates the minimum number of edit operations required to transform one code statement into another. A smaller edit distance indicates a higher similarity.
Hash values comparison~\cite{hummel2010hash} generates hash for each code statement through MD5 algorithm and then compares the hash similarity.
CodeBERT embedding comparison~\cite{sonnekalb2022codebert} uses CodeBERT to generate embedding representations for each code statement, then computes the cosine similarity between these embeddings.
FCDetection~\cite{FCDetector} generates AST and CFG representations as features. Word2vec and Graph2vec are used to embed these features. Then the fused feature vectors are input into a deep neural network model for classification of code clones.

\begin{table}[t]
    \centering
    \caption{Accuracy of Code Clone Detection Methods.}
    \label{tab:codeclone}

    \begin{tabular}{@{} llcc @{}} % Use booktabs-friendly specifiers
        \toprule
        \textbf{Methods} & \textbf{Level} & \textbf{Precision} & \textbf{Recall} \\ % Fixed typo
        \midrule
        Edit distance    & statement & 0.85 & 0.84 \\
        \midrule
        Hash values      & statement & 0.82 & 0.82 \\
        \midrule
        CodeBERT         & statement & 0.84 & 0.88 \\
        \midrule
        FCDetector       & function  & 0.89 & 0.93 \\
        \midrule
        \textsc{Vercation} & statement & \textbf{0.90} & \textbf{0.95} \\
        \bottomrule
    \end{tabular}
    
\end{table}

For Edit distance, Hash values, and CodeBERT embedding comparisons, we conducted independent threshold optimization. Similar to our threshold sensitivity analysis for Vercation, we evaluated each method across different threshold values from 0.1 to 1.0 with a step of 0.1, and selected the threshold that achieved the best F1 for that specific method. As FCDetector is a specialized code clone detection tool, we used its original pre-trained model and thresholds as specified in their work. The results presented in Table~\ref{tab:codeclone} represent each method's optimal performance with their respective optimized thresholds.
The results show that the AST-based code clone detection \hl{achieves} the best performance. 
Because the edit distance and hash value methods terminate the backtrack in cases of low textual similarity without considering code behavior, thus leads to a large number of false negatives. The results show that our method is robust against low syntactic similarity.
CodeBERT embedding comparison, while effective for semantic similarity, cannot capture fine-grained structural information. 
AST representations preserve the exact syntactic structure of code, allowing for more precise comparisons of code organization and logic flow. This structural analysis can identify clones that CodeBERT might miss, especially in cases where similar functionality is implemented with different coding patterns or variable names, which can be addressed by AST normalization.
Although FCDetector also extracts AST as a feature and normalizes variable names and constant values, it cannot identify function outline cases (as shown in Listing~\ref{semantic}). 
Moreover, the CFG features that FCDetector focuses on are not particularly effective when there are only a few vulnerable statements.

\subsubsection{Analysis of Refactoring Commits}
As motivated in Example 2 of Section~\ref{sec:motivating}, refactoring commits can significantly impact vulnerable version identification by introducing structural differences while preserving code behavior. To validate this motivation and evaluate the effectiveness of our code clone detection approach, we analyzed the commits encountered during our evaluation.
In the backtrack of our dataset, we conducted a total of 276 comparisons between pre-commit and post-commit. On average, it takes 3.73 comparisons to identify the \textit{vic} for each \textit{vfc}.
During the commit backtrack, cases with significant edit-distance differences ($\theta_1 < 0.9$) but minor AST differences ($\theta_2 \ge 0.8$) were classified as refactoring commits.
Out of these comparisons, 89 commits (32.2\%) were identified as refactoring commits, indicating that refactoring commits are a significant factor in backtracing \textit{vic}.
Furthermore, our analysis reveals that cases without any intermediate commit between the \textit{vic} and the \textit{vfc} are comparatively uncommon, constituting only 11\% of instances in our dataset.

\subsubsection{Threshold Sensitivity}
We used $\vartheta_1$ and $\vartheta_2$ to represent the threshold for edit-distance similarity and AST-based similarity, respectively. $\vartheta_3$ represents the \textit{similarity\_score} used for the backtrack termination condition. To measure the sensitivity of the thresholds, we incrementally increased $\vartheta_1$, $\vartheta_2$ and $\vartheta_3$  by 0.1 from 0 to 1, and evaluated the identification F1-score of the CVEs in ground-truth. 
Specifically, we first initialized $\vartheta_2$ and $\vartheta_3$ to 0.8, as high similarity thresholds generally provide better precision in code clone detection. Then we varied $\vartheta_1$ from 0.1 to 1.0 with a step of 0.1 and found the value of $\vartheta_1$ that achieved the highest F1-score (0.9).
With $\vartheta_1$ fixed at 0.9 and kept $\vartheta_3$ at 0.8, we varied $\vartheta_2$ from 0.1 to 1.0 and found the optimal value  (0.8). Finally, with $\vartheta_1$ = 0.9 and $\vartheta_2$ = 0.8, we varied $\vartheta_3$ from 0.1 to 1.0 and determined the optimal value of $\vartheta_3$ (0.7).
When the thresholds are greater than the optimal values, the precision is higher while the recall decreases. In contrast, the precision slightly decreases.

\begin{figure}[t]
	\centering
	\includegraphics[width=1\linewidth]{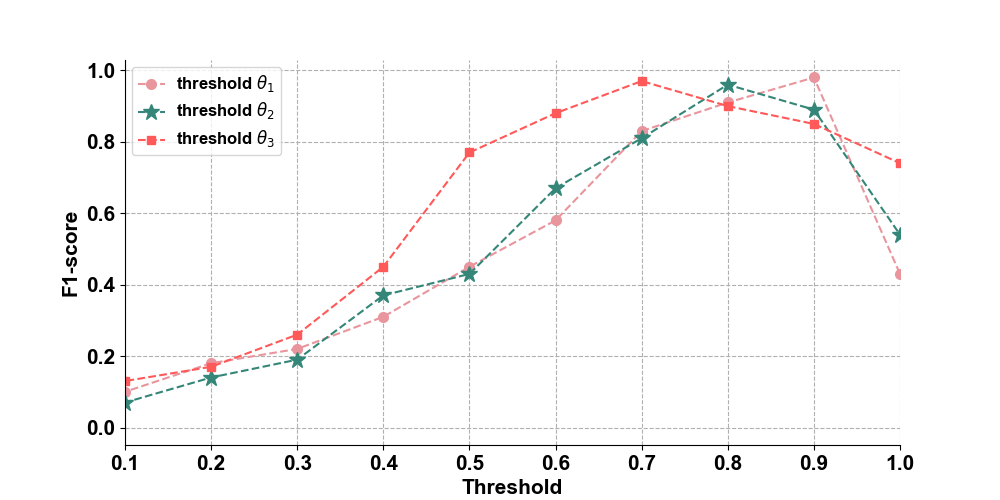}
	\caption{Threshold Sensitivity of \textsc{Vercation}.}
	\label{fig:threshold}
\end{figure}

\begin{center}
\fcolorbox{black}{gray!10}{\parbox{0.96\linewidth}{\textbf{RQ3}: Refactoring commits occupy a significant proportion of the code changes history. The AST-based similarity comparison in \textsc{Vercation} performs better in the vulnerability-introducing commit identification task for calculating code similarity, enhancing the overall effectiveness of \textsc{Vercation} in vulnerable versions detection.}}
\end{center}

\subsection{RQ4: Real-World Application}
We created a new vulnerability dataset consisting of 342 CVEs from the NVD. By developing a simple crawler, we obtained the corresponding CPE and CVE patches. We applied \textsc{Vercation} to the vulnerability dataset and found the vulnerable versions of 202 CVEs are incorrect. Among them, the CPE for 134 CVEs is incomplete, with 108 CVEs only reporting a single vulnerable version, significantly increasing the risk of vulnerability propagation. 
Furthermore, the CPE for 68 CVEs is affected by the overinclusion problem, with 56 CVEs considered all versions before the OSS version mentioned in the vulnerability report as susceptible to attack, without conducting a detailed analysis. 
We have submitted reports to the NVD detailing the CVEs with incorrect affected versions information we identified. The NVD team's response said they plan to address these inaccuracies as part of their ongoing efforts to enhance the quality of their vulnerability information.
More importantly, we have discovered 4 CVEs that do not provide information about the vulnerable version.

% \noindent\textbf{Vulnerability Severity.}
To understand the practical impact of accurately identifying vulnerable versions, we analyzed the severity of 134 vulnerabilities with incomplete version information in the NVD dataset.
The analysis shows that CVEs with CVSS Score of 4.0-6.9 (Medium) and 7.0-8.9 (High) accounted for the most, at 49.7\% and 49.1\%. 
Notably, we identified 2 critical vulnerabilities (CVSS Score 9.0-10.0) in \textit{FFmpeg} and \textit{php}. 
These findings highlight that in many high-severity vulnerabilities, NVD does not report the correct vulnerable version. This inaccuracy has serious implications, as downstream software manufacturers may not recognize the need to update their vulnerable dependencies in time, exposing their systems to significant security risks. Vercation's high accuracy in identifying vulnerable versions therefore helps organizations improve the accuracy of vulnerability reports.
%The result indicates that the impact of OSS vulnerabilities is severe, we need to pay more attention to OSS vulnerabilities.

%In particular, we found that there are only 2 vulnerabilities with a CVSS score of Low (CVSS Score in 0.1-3.9), indicating that the impact of OSS vulnerabilities is severe, we need to pay more attention to OSS vulnerabilities.
\begin{center}
\fcolorbox{black}{gray!10}{\parbox{0.96\linewidth}{
\textbf{RQ4}: %The NVD's vulnerable version information was found to be incomplete and inaccurate. \textsc{Vercation} can serve well as a supplement to NVD information. Additionally, We disclosed the time delay in fixing and publishing OSS vulnerabilities.
%Applying Vercation to real-world CVEs revealed significant practical impact. It identified 134 CVEs (38.1\%) with incomplete or inaccurate CPE information in NVD reports. Analysis of the top 10 OSS projects showed substantial delays in vulnerability resolution, with some cases lasting up to ten years. This demonstrates Vercation's potential to enhance vulnerability management in real-world scenarios by providing more accurate version information and insights into vulnerability lifecycles.
Applying Vercation to real-world CVEs demonstrated practical impact in two key findings: it identified 202 CVEs (59.1\%) with incomplete or overincluded versions in NVD reports. Notably, nearly half of the vulnerabilities with incorrect versions were of high or critical severity, emphasizing the importance of accurate version identification.
% \zt{The numbers here need to be updated too}
}}
\end{center}

\section{Discussion}
\subsection{Performance and Cost Analysis}
In terms of performance efficiency, \textsc{Vercation} analyzes each vulnerability in an average of 28.61 seconds. This includes the time for program slicing, LLM processing, and AST-level code clone detection. 
Compared to existing approaches: V-SZZ takes 10.25 seconds on average, V0Finder requires approximately 22 seconds per CVE, \hl{Vision needs 1,094.12 seconds per vulnerability}, and V1SCAN can detect vulnerabilities within 20 seconds for 99\% of projects. While VERJava performs faster at 0.71 seconds per CVE, it achieves much lower recall in version identification. Compared to other methods, Vercation achieves higher accuracy without significant time overhead.

\hl{The best LLM DeepSeek-V3 in our evaluation, whose API pricing is \$0.30 per million tokens for input and \$1.10 per million tokens for output. Based on our experimental setup with 10 independent runs for self-consistency, the average LLM API cost per CVE is only \$0.0034. This cost includes: input tokens for CVE descriptions, dangerous flows, and few-shot examples, output tokens for logic explanation, and vulnerable statements. The low cost allows the LLM’s application in practical security.}

\subsection{Comparison with Dynamic Analysis}
\hl{While dynamic analysis through PoC testing provides definitive evidence of vulnerability existence, several limitations render it insufficient for comprehensive and efficient identification of vulnerable versions at scale. Our analysis of CVEs published in the NVD from 1999 to 2025 reveals that only 24\% of CVEs provide publicly available exploit links, and these exploit reports do not guarantee successful reproduction across different software versions.}

\hl{The portability of PoCs across different software versions presents another critical challenge for dynamic analysis approaches. Previous studies have demonstrated that PoC reproducibility across different environments and versions is a significant technical hurdle}~\cite{incomplete_mu2018understanding, dai2021facilitating}. \hl{During the construction of our ground-truth dataset, we encountered these challenges firsthand when attempting to validate vulnerabilities through PoC execution. Of the available PoCs in our dataset, only 55\% could be successfully executed across multiple versions without modification. The remaining cases failed to trigger the vulnerability due to environmental dependencies, compilation differences, or version-specific behavioral changes, which required substantial manual code-level analysis for verification.} 

\hl{The temporal cost of dynamic analysis further limits its practical applicability for large-scale vulnerable version identification. Our experience during dataset construction revealed that each CVE required an average of 3 hours for environment setup, dependency resolution, and PoC execution across different versions, excluding the time needed for manual verification when PoCs failed. For our complete dataset of 122 CVEs spanning 1,111 versions, comprehensive dynamic testing would require approximately 30 days compared to 28.61 seconds per CVE for our static approach. This represents a 742x time difference in favor of static analysis, making dynamic approaches impractical for real-world deployment scenarios where rapid vulnerability assessment is crucial.}

%\noindent\textbf{Branch Fixing Oversights.} In some cases, we observe that security fixes are promptly incorporated into the latest version branches of OSS. However, rectifying oversights frequently transpires during the backporting of fixes to older release branches. In such scenarios, projects not utilizing the latest branch version may encounter a deficiency in receiving security updates. Simultaneously, the latest versions updated in older branches remain exposed to risks.

\subsection{Threats to Validity}
\subsubsection{Internal Validity}
% \vspace{4px}
\noindent\textbf{Static–analysis precision.}
\textsc{Vercation} relies on the Joern parser to build code-property graphs that combine an AST, control-flow graph, and data-dependency graph. 
Joern struggles with certain C/C\texttt{++} constructs, notably complex pointer arithmetic, overloaded constructors, and indirect calls, so some data flows and constructor edges are missed.  
\hl{Our AST expansion step further inherits the classic limitations of method outlining: it cannot always disambiguate targets reachable via function pointers, callbacks, or C\texttt{++} virtual dispatch, which static analysis alone cannot resolve definitively} \cite{sui2016svf,christakis2016developers}.  
\hl{These imprecisions may hide or mislocate refactoring patterns.}

% \noindent\textbf{Limitation of Joern Parser.}
% \textsc{Vercation} utilizes the Joern parser to generate the code property graph, containing abstract syntax tree, control flow graph, and data dependency graph. However, 
% we encountered some challenges with certain C/C++ constructs that Joern did not handle well, impeding its ability to track data flow and missing essential constructors correctly. Additionally, Joern may face difficulties in outputting data dependencies related to pointers, particularly in languages like C/C++ where pointers are heavily used.
% To counter this, we leverage the power of LLM.

% \vspace{4px}
% \noindent\hl{\textbf{Limitation of Function Expansion.} Our AST expansion technique for handling method outlining has inherent limitations when dealing with indirect function calls, such as function pointers, callback functions, and virtual function dispatch in C++. In such cases, static analysis is hard to resolve the target function definitively}~\cite{sui2016svf,christakis2016developers}. \hl{This limitation may affect the detection of some refactoring patterns involving indirect calls.}

\vspace{4px}
\noindent\textbf{Inter-procedural dependencies.}
\hl{In multi-function patches, \textsc{Vercation} currently analyses each affected function independently and then merges the results.  
Although effective for most cases, this strategy may overlook vulnerability patterns that hinge on data/control flows spanning several functions.}

\vspace{4px}
\noindent\textbf{LLM variability and reproducibility.}
As LLMs continue to evolve rapidly, the performance of \textsc{Vercation} might change with newer versions of these models. This could affect the long-term consistency of our results. Additionally, LLMs may have biases or limitations in their training data that could influence their ability to understand certain types of vulnerabilities or code structures. To mitigate this, we have provided detailed information about the LLM versions and prompts used in our study, but future research may need to account for potential variations in LLM capabilities and performance as these models continue to develop.

\subsubsection{External Validity}
\noindent\textbf{Language generalisability.} \textsc{Vercation}'s core methodology is language-agnostic and can be adapted to other programming languages. The program slicing tool Joern inherently supports multiple languages, including Java, Python, JavaScript, and PHP. The main adaptations required would be updating the sensitive function table for language-specific vulnerability patterns and modifying AST generation rules according to the target language's syntax and semantics. 
Future work could involve extending \textsc{Vercation} to support a broader range of programming languages and verifying its effectiveness across different language paradigms.

\vspace{4px}
\noindent\textbf{Repository branching complexity.}
\hl{While our approach handles common merge}~\cite{rios2022unifying} \hl{and branching patterns effectively, complex branching scenarios, such as when the same vulnerable code is independently modified in multiple parallel feature branches before being merged (creating diamond-shaped merge patterns), may require more sophisticated graph traversal algorithms to accurately determine the earliest vulnerability introduction point}~\cite{fan2019branch, rosa2021branch}.

% \subsection{Limitation and Future Work}
% \noindent\textbf{Limitation of Assumptions.} 
% \textsc{Vercation} operates under certain assumptions that may limit its applicability. Firstly, it is tailored for C/C++ projects, yet its methodology can be applied to other programming languages. With an accurate exploration of common vulnerability types in diverse languages, \textsc{Vercation} could identify vulnerable versions in those languages.
% Secondly, 
% \hl{\textsc{Vercation} processes multi-function patches by analyzing each affected function independently and combining the results. While this approach handles most vulnerability scenarios, the inter-procedural dependencies may not be fully captured in the current implementation. Future work will explore integrating inter-procedural analysis techniques to enhance the detection of such cross-functional vulnerability patterns.}

% \vspace{4px}
% \noindent\hl{\textbf{Branching Scenarios.} While our approach handles common merge}~\cite{rios2022unifying} \hl{and branching patterns effectively, complex branching scenarios, such as when the same vulnerable code is independently modified in multiple parallel feature branches before being merged (creating diamond-shaped merge patterns), may require more sophisticated graph traversal algorithms to accurately determine the earliest vulnerability introduction point}~\cite{fan2019branch, rosa2021branch}.

\section{Related work}
\subsection{Vulnerability Fixing Commit Analysis}
%Soto et al.~\cite{patch_soto2016deeper} conduct a large-scale study of bug-fixing commits in Java projects, guiding high-quality automatic software repair to target. Perl et al.~\cite{patch_perl2015vccfinder} create a CVE-linked commit database and an SVM-based suspicious commit classifier to find potentially dangerous code in code repositories. 
Several studies have focused on analyzing vulnerability fixing commit (VFC) to extract vulnerability features, which can be used for various purposes such as vulnerability detection, classification, and affected version identification.
VulPecker~\cite{li2016vulpecker} extracts code patterns from VFC to detect similar vulnerabilities in other software versions.
SySeVR~\cite{li2021sysevr} uses a systematic approach to extract syntax- and semantic-based vulnerability features from patches.
VulDeePecker~\cite{li2018vuldeepecker} leverages deep learning techniques to learn vulnerability patterns from code gadgets extracted from vulnerability fixing commits.
%VulCNN~\cite{wu2022vulcnn} uses convolutional neural networks to learn vulnerability patterns from code changes in VFC.
Most existing VFC analysis methods rely heavily on predefined rules or patterns, which may not capture the full context of vulnerabilities. %They also often struggle with handling complex code changes and may miss subtle vulnerability indicators.
\textsc{Vercation} incorporates VFC analysis by leveraging LLMs to understand the context of code changes in fixing commits, potentially capturing more nuanced vulnerability features.

\subsection{Vulnerable Version Detection for OSS}
Several approaches have been proposed for vulnerable version identification. V-SZZ~\cite{bao2022v} uses the SZZ algorithm to backtrack \textit{vfc}s and identify vulnerability-introducing changes. MVP~\cite{xiao2020mvp} employs program slicing to extract vulnerability and patch signatures, then uses these to identify potentially vulnerable functions. %AFV~\cite{shi2022precise} introduces a vulnerability-centric approach for precise analysis of affected versions in web vulnerabilities.
V0Finder~\cite{woo2021v0finder} generates fingerprints for vulnerable functions and uses a clone-based technique to detect vulnerable versions across different software releases. VCCFinder~\cite{perl2015vccfinder} uses machine learning techniques to identify vulnerability-contributing commits. VUDDY~\cite{kim2017vuddy} proposes a scalable approach for vulnerable code clone detection in large-scale code bases.
Existing approaches often rely on predefined patterns and syntactic-level analysis, leading to imprecision in vulnerability characterization and code clone detection.
\textsc{Vercation} combines static analysis with LLM for code understanding and introduces AST-based code clone detection to address the code structure modification problem.

\subsection{Code Refactoring Detection}
Some researchers conducted studies on the characteristics of code refactoring not altering the code behavior of the code and have proposed detection methods.
MLRefScanner~\cite{noei2024refactor} detects refactoring commits in Python projects by analyzing commit history and extracting features that represent refactoring activities without altering code behavior.
Abid et al.~\cite{abid2020refactor} perform a study outlining different levels of refactoring from code level to architecture and discuss the importance of maintaining behavior during refactoring.
Eman et al.~\cite{alomar2021refactoring} focus on how refactoring is integrated into the code review process while ensuring that the original software behavior is preserved.
Existing code refactoring detection methods primarily focus on high-level changes like renaming and method/class moves. Our approach, however, targets statement-level changes through fine-grained AST expansion and normalization, including inline function expansion during AST generation. %This level of detail in internal code structure analysis is a novel aspect not previously considered.
%Some state-of-the-art techniques utilize semantic analysis to conduct code clone detection~\cite{clone_semantic_yu2023graph,semantic_fang2020functional,semantic_hu2022treecen,semantic_svacina2020semantic,semantic_wu2020scdetector}. Yu et al.~\cite{clone_semantic_yu2023graph} propose a novel graph-based code semantics learning method to capture critical information at token, statement, edge, and graph levels. 
%Jan et al,~\cite{semantic_svacina2020semantic} identify semantic code clones in enterprise frameworks by using control flow graphs and applying various proprietary similarity functions to compare enterprise targeted metadata for each pair of CFGs. 
%SCDetector~\cite{semantic_wu2020scdetector} feeds tokens with graph detail into a Siamese architecture neural network to train a code clone detector.
%Treecen~\cite{semantic_hu2022treecen} treats the AST graph as a social network and adopt centrality analysis on each node to maintain the tree details. By this, the original complex tree can be converted into a 72-dimensional vector while containing comprehensive structural information of the AST. Finally, these vectors are fed into a machine learning model to train a detector and use it to find code clones.
%Fang et al.~\cite{semantic_fang2020functional} propose a joint code representation that applies fusion embedding techniques to learn hidden syntactic and semantic features of source codes, then train a supervised deep learning model to detect functional code clone.

\subsection{Code Clone Detection}
Some SOTA techniques utilize static analysis to conduct code clone detection.
%Jan et al,~\cite{semantic_svacina2020semantic} identify code clones in enterprise frameworks by using control flow graphs and applying various proprietary similarity functions to compare enterprise-targeted metadata for each pair of CFGs. 
SCDetector~\cite{semantic_wu2020scdetector} feeds tokens with graph detail into a Siamese architecture neural network to train a code clone detector.
%Treecen~\cite{semantic_hu2022treecen} treats the AST graph as a social network and converts it into a 72-dimensional vector while containing comprehensive structural information about the AST. Finally, these vectors are fed into a machine-learning model to train a detector and use it to find code clones.
Fang et al.~\cite{semantic_fang2020functional} propose a joint code representation that applies fusion embedding techniques to learn hidden features of source codes, then train a supervised deep learning model to detect functional code clones.
FCDetection~\cite{FCDetector} generates AST and CFG representations and uses Word2vec to embed these features into vectors. Then the fused feature vectors are input into a deep neural network model for classification of code clones.

\section{Conclusion and Future Work}
This paper proposed \textsc{Vercation}, an approach designed for vulnerable version identification of open-source C/C\texttt{++} software. %The core idea involves a symbiotic combination of static analysis, LLM, and AST-level code clone detection.
Our approach introduces two key innovations: leveraging LLMs to enhance the extraction of vulnerability-related statements and employing code clone detection based on expanded and normalized ASTs.
Experimental results on our dataset of 122 CVEs across 12 popular open-source projects validate that \textsc{Vercation} surpasses existing techniques for vulnerable version identification, demonstrating notable improvements in both precision and recall.
\hl{\textsc{Vercation}'s novelty lies not in individual techniques but in the systematic integration of semantic analysis and methodological innovations in LLM application. The combination creates a qualitatively different approach that advances the state-of-the-art in both accuracy and practical applicability, opening new research directions for AI-assisted software security analysis.}

\hl{In the future, we plan to empirically validate \textsc{Vercation} on a broader set of languages (Java, Python, JavaScript, and PHP) to confirm its effectiveness. Furthermore, we are interested in exploring integrating inter-procedural analysis techniques to enhance the detection of cross-functional vulnerability patterns.}
%Equipped with the discovery from \textsc{Vercation}, developers can effectively address potential threats arising from cross-version vulnerabilities in OSS, thereby enhancing the overall security of the software ecosystem.

\section*{Acknowledgement}
\hl{We thank the anonymous reviewers for their insightful comments. This research is supported by Beijing Natural Science Foundation (Grant No. L234033), the National Research Foundation, Singapore, and the Smart Nation Group under the Smart Nation Group’s Translational R\&D Grant (Award No. TRANS2023-TGC02). Any opinions, findings and conclusions or recommendations expressed in this material are those of the author(s) and do not reflect the views of National Research Foundation, Singapore or the Smart Nation Group.}

\balance
\bibliographystyle{IEEEtran}
\bibliography{main}

\end{document}